\documentclass[12pt,preprint]{emulateapj}
\usepackage{natbib}

\newcommand{\be}{\begin{equation}}
\newcommand{\ee}{\end{equation}}
\newcommand{\ba}{\begin{eqnarray}}
\newcommand{\ea}{\end{eqnarray}}

\begin{document}

\title{Intensity Mapping of the [CII] Fine Structure Line during the Epoch of Reionization}

\author{Yan Gong$^1$, Asantha Cooray$^1$, Marta Silva$^2$, Mario G. Santos$^2$, James Bock$^{3,4}$, C. Matt Bradford$^{3,4}$, Michael Zemcov$^{3,4}$}

\affil{$^1$Department of Physics \& Astronomy, University of California, Irvine, CA 92697}
\affil{$^2$CENTRA, Instituto Superior T\'ecnico, Technical University of Lisbon, Lisboa 1049-001, Portugal}
\affil{$^3$California Institute of Technology, 1200 E. California Blvd., Pasadena, CA 91125, USA}
\affil{$^4$Jet Propulsion Laboratory, 4800 Oak Grove Drive, Pasadena, CA 91109, USA}

\begin{abstract}

The atomic CII fine-structure line is one of the brightest lines in a typical star-forming galaxy 
spectrum with a luminosity $\sim 0.1$\% to 1\% of the bolometric luminosity. 
It is potentially a reliable tracer of the dense gas  distribution at high redshifts and could
provide an additional probe to the era of reionization. By taking into account of the spontaneous, 
stimulated and collisional emission of the CII line, we calculate
the spin temperature and the mean intensity as a function of the redshift. When averaged
over a cosmologically large volume,  we find that the CII emission from ionized carbon in
individual galaxies is larger than the signal generated by carbon in
the intergalactic medium (IGM). Assuming that the CII luminosity is proportional to the carbon mass in dark matter halos, 
we also compute the power spectrum of the CII line intensity at various redshifts. 
In order to avoid the contamination from CO rotational lines at low redshift
when targeting a CII survey at high redshifts, we propose the cross-correlation of CII and 21-cm line emission
from high redshifts. To explore the detectability of the CII signal from reionization, we also evaluate the expected errors on 
the CII power spectrum and CII-21 cm cross power spectrum based on the design of the future milimeter surveys.
We note that the CII-21 cm cross power spectrum contains interesting features that captures physics during reionization, including the
ionized bubble sizes and the mean ionization fraction, which are challenging to measure from 21-cm data alone.
We propose an instrumental concept for the reionization CII experiment  targeting the frequency range of $\sim$ 200 to 300 GHz
with 1, 3 and 10 meter apertures and a bolometric spectrometer array with 64 independent spectral pixels with about 20,000 bolometers.

\end{abstract}


\maketitle

\section{Introduction}

Carbon is one of the most abundant elements in the Universe and it is
 singly ionized (CII) at $11.26$ eV, an ionization energy that is less than that of the
hydrogen. With a splitting of  the fine-structure level at $91\ \rm K$ CII is easily excited resulting
in a line emission at $157.7\ \rm \mu m$ through the $^2P_{3/2}\to ^2P_{1/2}$ transition.
It is now well established that this line provides a major cooling mechanism
for the neutral interstellar medium (ISM) \citep{Dalgarno72, Tielens85,Wolfire95,Lehner04}. 
It is present in
multiple phases of the ISM in the Galaxy \cite{Wright91} including the most diffuse regions \cite{Bock93}
and the line emission has been detected from
the photo dissociation regions (PDRs) of star-forming galaxies \citep{Boselli02,DeLooze11,Nagamine06,Stacey10} and, in some cases, even in $z > 6$ Sloan
quasars \citep{Walter09}.

The CII line is generally the  brightest emission line in star-forming galaxy
spectra and contributes to about 0.1\% to 1\% of the total far-infrared (FIR)
luminosity \citep{Crawford85,Stacey91}. 
Since carbon is naturally produced in stars, CII emission is then expected
to be a good tracer of the gas distribution in galaxies. Even if the angular 
resolution to resolve the CII emission from individual galaxies is not available, 
the brightness variations of the CII line intensity can be used to map the 
underlying distribution of galaxies and dark matter
\citep{Basu04,Visbal10,Gong11}. 

Here we propose CII intensity mapping as an  alternative avenue to probe the era of reionization, including the transition 
from primordial galaxies with PopIII stars alone to star-formation in  the second-generation of galaxies
with an ISM polluted by  metals. CII intensity mapping complements attempts to study reionization 
with low-frequency radio experiments that probes the 21-cm spin-flip line from
neutral hydrogen. While those experiments are sensitive to the neutral gas 
distribution dominated by the IGM, CII will probe the
onset of star formation and metal production in $z \sim 6$ to 8 galaxies.

Recently it has also been proposed to use rotational lines of CO molecule to probe reionization (e.g., Gong et al. 2011; Carilli 2011; Lidz et al. 2011).
CO studies have the advantage that redshift identification is facilitated by multiple $J$ transition lines, while with CII some confusion
could result in the line identification with other atomic fine-structure and molecular lines at sub-mm and mm wavelengths.
In comparison, and based on limited observational data, CII emission is expected to brighter in low mass galaxies, compared to the case of CO
luminosity. It is unclear if the galaxies present during reionization are analogous to low-redshift dwarf galaxies or the local ultra-luminous infrared
galaxy population. At high redshifts, we expect most of the carbon to be in CII  rather than 
CO, since the high-redshift galaxies may not host significant dust columns
required to shield CO from dissociating UV photos. Given that a CO experiment to study reionization will involve an experiment at mid radio frequencies
of 15 to 30 GHz, while CII will involve high frequencies, the two methods will be affected by different systematics and foregrounds. Thus, a combined
approach involving multiple probes of reionization, 21-cm, CO and CII, would provide the best avenue to study the $z > 6$ universe.

In this paper we present an analytical calculation to predict the CII intensity
intensity as a function of the redshift by considering spontaneous, stimulated and collisional emission processes
\citep{Suginohara99,Basu04}. This intensity changes the
brightness spectrum of the cosmic microwave background (CMB) at the frequency corresponding to the
CII line (Basu et al. 2004).  In this paper, we focus
on the CII flux from individual galaxies where the matter density is
high and the collisional emission is the dominant process. As a check on our analytical calculations, 
we also consider results derived from numerical simulations to establish the CII intensity.
The two approaches are generally consistent. 
We then consider the measurement of the CII intensity fluctuations resulting from the clustering of 
the galaxy distribution and sources that are present during and towards the end of reionization at $z \sim 6-8$.

Experimentally, there are several challenges that one must overcome before the CII intensity fluctuations from
high redshifts can be reliably established. First, higher J transitions of CO from dense molecular gas at
lower redshifts contaminate the CII line intensity measurements. In particular, one must account for all of CO(2-1) to CO(13-12) emission lines
from individual narrow redshift ranges in the foreground between 0 and 6 when probing CII fluctuations at $z > 6$.
To the extent to which a variety of existing predictions on the CO intensity can be trusted \citep{Gong11,Lidzetal11,Carilli11},
we generally find that the contamination is mostly below the level of the CII signal.
Extending previous studies \citep{Basu04,Visbal10,Gong11,Lidzetal11}, we propose the cross-correlation of CII line intensity mapping
and 21-cm fluctuations as a way to improve studies related to the epoch of reionization.
To evaluate the detectability of the CII signal, we calculate the errors on the CII power spectra and CII-21cm cross correlation, respectively,
based on the design of potential (sub-)milimeter  surveys for the CII emission. For 21-cm, we consider the first generation experiment
 LOw Frequency ARray\footnote{http://www.lofar.org/} (LOFAR) survey as well as the improvement expected from a second generation experiment
like the low-frequency extension to the Square Kilometer Array (SKA).

The paper is organized as follows: in the next section, we derive
the formulas to calculate the spin temperature
of the CII line in the ISM of galaxies and the IGM. In Section~3, we calculate
the mean CII intensity analytically and compare it with results derived from a simulation.
We show the CII power spectrum in Section~4.
In Section~5, we discuss the low-redshift contamination from CO emission lines for CII intensity mapping at $z > 6$
and in Section~6 we propose a  cross-correlation between CII and 21-cm line intensity measurements over the overlapping redshift ranges as a way to
both distinguish the CII signal from the CO emission and to improve overall constraints on reionization physics.
In Section~7 we outline the experimental parameters of a CII spectral mapping experiment designed to probe $z \sim 6$ to 8 CII intensity fluctuations
and discuss the detectability of the CII power spectrum and the CII and 21-cm cross-correlation when combined with LOFAR and SKA.
We conclude with a summary of our results in Section~8. 
Throughout this paper, we assume the flat $\rm \Lambda CDM$ model with
$\Omega_{m} = 0.27$, $\Omega_b=0.046$, $\sigma_8=0.81$, $n_s=0.96$
and $h=0.71$ \citep{WMAP7}. 

\section{The derivation of the spin temperature of the CII emission}

The CII line intensity can be generated from carbon present in both the interstellar medium (ISM) of individual galaxies and the diffuse
intergalactic medium (IGM) in between galaxies.  In the ISM of galaxies, CII is expected to be in thermodynamic equilibrium with resonant
scattering emission off of CMB photons and the collisional emission induced by gas in the galaxy. If the number density of gas particles $n_{\rm gas}$ is 
greater than a critical value $n_{\rm cr}$, the collisional excitation 
and de-excitation rate would exceed that of the radiative (spontaneous and stimulated) processes. 

Since the ionization potential of carbon is  $11.26\ \rm eV$, below 13.6 eV of hydrogen ionization,
the surrounding gas of CII ions can be either neutral hydrogen or electrons. However, 
because the critical number density of electrons to trigger collisional excitation
$n_{\rm e}^{\rm cr}$ is less than $100\ \rm cm^{-3}$ while the critical 
number density of neutral hydrogen for collisional excitation
$n_{\rm H}^{\rm cr}$ is about $10^3$ to $10^4\ \rm cm^{-3}$ 
\citep{Malhotra01}, the electrons can collide with CII ions more 
frequently than with HI, especially in ionized gas 
\citep{Lehner04,Suginohara99}. For simplicity and not losing generality,
we assume that the ISM is mostly ionized in individual galaxies. Then the CII line emission would mainly be excited by electrons  in the ISM
\citep{Suginohara99}. 

On the other hand, in the diffuse IGM in between galaxies CII line emission will be
 mainly due to radiative processes such as spontaneous emission,  stimulated emission due to collisions with CMB photons, and a UV pumping effect 
similar to the Ly-$\alpha$ coupling for the 21-cm emission \citep{Wouthuysen52,Field58,Hernandez06}.
We will focus on the CII emission from the ISM of galaxies first and then
consider the signal from the IGM. The latter is found to be negligible.

\subsection{The CII spin temperature in the ISM of galaxies}

In the ISM of galaxies, the ratio of thermal equilibrium 
population of the upper level $^2P_{3/2}$ and lower level $^2P_{1/2}$
of CII fine structure line can be found by solving the statistical balance 
equation
\ba \label{eq:n_ratio}
\frac{n_{\rm u}}{n_{\rm l}} &=& \frac{B_{\rm lu}I_{\nu}+n_{\rm e}C_{\rm lu}}{B_{\rm ul}I_{\nu}+A_{\rm ul}+n_{\rm e}C_{\rm ul}}\\  \nonumber
                &\equiv& \frac{g_{\rm u}}{g_{\rm l}}{\rm exp}[-T_{\star,\rm ul}/T_{S,\rm ul}]. 
\ea
Here $A_{\rm ul}=2.36\times 10^{-6}\ {\rm s^{-1}}$ is the spontaneous 
emission Einstein coefficient \citep{Suginohara99}, $B_{\rm ul}$ and $B_{\rm lu}$ are the 
stimulated emission and absorption coefficients, respectively, 
$I_{\nu} \equiv B[T_{\rm CMB}(z)]$ is the intensity of CMB at $\nu_{\rm ul}$, 
$n_{\rm e}$ is the number density of electrons, and $C_{\rm lu}$ and 
$C_{\rm ul}$ are the excitation and de-excitation collisional rates
(in $\rm cm^3s^{-1}$), respectively. Note that the UV pumping effect is neglected 
here and, as we discuss later, it should not affect the result unless the UV intensity 
inside the galaxy ($I_{\rm UV}^{\rm gal}$) is higher than $10^{-15}\ \rm erg/s/cm^2/Hz/sr$, which
is about $10^6$ times greater than the UV background \citep{Giallongo97,Haiman00}.
We note that even if the radiative coupling from UV pumping made a small 
contribution to the spin temperature it would be in the same direction as the collisional coupling 
since the UV color temperature follows the gas temperature.

The second line of Eq.\ref{eq:n_ratio} defines the excitation or 
spin temperature $T_{S, \rm ul}$ of the CII line. The statistical weights are $g_{\rm u}=4$ and 
$g_{\rm l}=2$, and  $T_{\star, \rm ul} \equiv h\nu_{\rm ul}/k_B \simeq 91 \rm K$ is the 
equivalent temperature of the level transition.
The excitation collisional rate $C_{\rm lu}$ can be written as
\citep{Spitzer78,Osterbrock89,Tayal08}
\be
C_{\rm lu} = \frac{8.629\times 10^{-6}}{g_{\rm l}\sqrt{T_{\rm k}^{\rm e}}}\gamma_{\rm lu}(T_{\rm k}^{\rm e}){\rm exp}\bigg( \frac{-T_{\star,\rm lu}}{T_{\rm k}^{\rm e}}\bigg), \label{eq:C_lu}
\ee
where $T_{\rm k}^{\rm e}$ is kinetic temperature of the electron and
$\gamma_{\rm lu}$ is the effective collision strength, a dimensionless quantity.

\begin{figure}[htb]
\includegraphics[scale = 0.43]{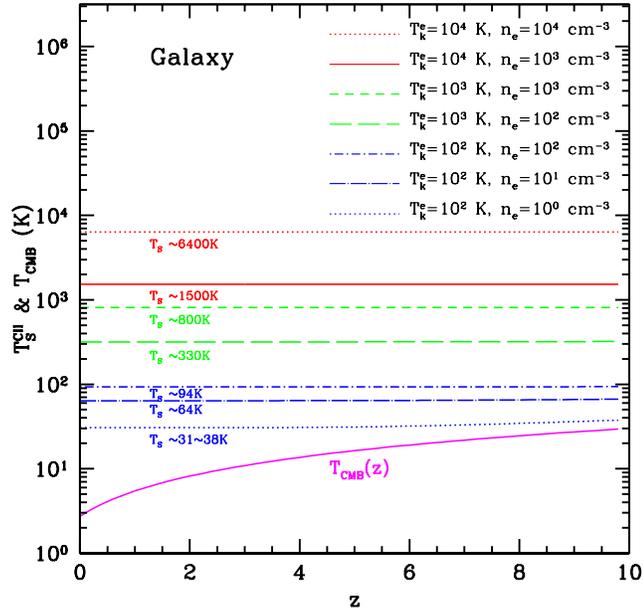} 
\caption{\label{fig:T_s_gal} The CII spin temperature $T_S^{\rm CII}$ and the
$T_{\rm CMB}$ in the ISM of galaxies as a function of the redshift. To capture different ISM conditions within a large sample
of galaxies we select several pairs of the electron kinetic temperature
$T_{\rm k}^{\rm e}$ and the number density $n_{\rm e}$ for this calculation.
We find that the CII spin temperature is almost constant with
redshift, which indicates that the collisional process
is dominant in the ISM of galaxies when compared to the radiative
processes.}
\end{figure}

To derive the spin temperature $T_S$, we can use Einstein relations
$g_{\rm l}B_{\rm lu}=g_{\rm u}B_{\rm ul}$ and 
$A_{\rm ul}=(2h\nu^3/c^2)B_{\rm ul}$
to convert both stimulated emission coefficient $B_{\rm ul}$ and absorption coefficient 
$B_{\rm lu}$ in terms of the spontaneous emission coefficient $A_{\rm ul}$. 
Also, using the collisional balance relation 
\be
\frac{C_{\rm lu}}{C_{\rm ul}}=\frac{g_{\rm u}}{g_{\rm l}}{\rm exp}[-T_{\star,\rm ul}/T_{\rm k}^{\rm e}], \label{eq:C_C}
\ee
we can finally write
\be
\frac{T_{\star,\rm ul}}{T_{S,\rm ul}} = log\Bigg\{ \frac{A_{\rm ul}[1+(I_{\nu}c^2/2h\nu^3)_{\nu_{\rm ul}}]+n_{\rm e}C_{\rm ul}}{A_{\rm ul}(I_{\nu}c^2/2h\nu^3)_{\nu_{\rm ul}}+n_{\rm e}C_{\rm ul}{\rm exp}(-T_{\star,\rm ul}/T_{\rm k}^{\rm e})} \Bigg\}. \label{eq:T_s}
\ee

Note that the de-excitation 
collisional rate $C_{\rm ul}$ is dependent on the $T_{\rm k}^{\rm e}$,
and using Eq.~\ref{eq:C_lu} and Eq.~\ref{eq:C_C},
we can calculate the de-excitation collisional rate $C_{\rm ul}$ for 
a fixed value of the electron kinetic temperature.
Here we adopt the values of $\gamma_{\rm lu}$ which are calculated by the
R-matrix in Keenan et al. (1986), and they find $\gamma_{\rm lu}=
1.58$, $1.60$ and $2.11$ at $T_{\rm k}^{\rm e}=10^2$, $10^3$ and 
$10^4\ \rm K$ respectively. 
Putting all values together, we find $C_{\rm ul}=3.41\times 10^{-7}$, $1.09\times 10^{-7}$ and 
$4.55\times 10^{-8}\ \rm cm^3s^{-1}$ for 
$T_{\rm k}^{\rm e}=10^2$, $10^3$ and $10^4\ \rm K$, respectively.
This is well consistent with the de-excitation rate of $4.6\times 10^{-8}\ \rm cm^3s^{-1}$ at 
$10^4\ \rm K$ given in \cite{Basu04}.

The deviation of the CII spin temperature $T_{S, \rm ul}$ relative to $T_{\rm CMB}$ as a function of redshift
for the electron dominated ISM of galaxies is shown in Fig.~\ref{fig:T_s_gal}.
We choose several values of the electron kinetic temperature 
$T_{\rm k}^{\rm e}$ and the number density $n_{\rm e}$ to plot the
spin temperature $T_S$ as a function of the redshift.
We note that although the mean number density of electrons 
$n_{\rm e}^{\rm mean}$ can be very small ($<10^{-2}\ \rm cm^{-3}$)
in a halo, the significant CII emission in galaxies comes from dense gas clumps which have a much higher $n_e$ \citep{Suginohara99}. Due to gas clumping the local
$n_{\rm e}$ can be much greater than $n_{\rm e}^{\rm cr}$ even though
$n_{\rm e}^{\rm mean}<n_{\rm e}^{\rm cr}$. Thus we choose to assume several high $n_{\rm e}$ values in our plots, but also show the case with
two low values of $n_{\rm e}=1$ and $10\ \rm cm^{-3}$. These values are less than $n_{\rm e}^{\rm cr}$.
We find the $T_S$ is almost constant and always greater than 
$T_{\rm CMB}$ for $0<z<10$ in all these cases. In Eq.~\ref{eq:T_s}, it is easy to
find that the $T_S$ depends on the relative strength of the
radiative (spontaneous and stimulated) and collisional processes.
If the spontaneous and stimulated emission are dominant, we have
$T_S\sim T_{\rm CMB}$, while $T_S\sim T_{\rm k}^{\rm e}$ if collisions
are dominant. Given a fixed number density $n_{\rm e}$, the only variable that depends on redshift $z$ in 
Eq.~\ref{eq:T_s} is $I_{\nu}(z)$, but the spin temperature is not strongly sensitive to it.
This implies that the collisional process is dominant in the ISM of galaxies  when compared to the
resonant scattering off of CMB photons. As we discuss next, this result is not true for 
the emission of the CII line in the diffuse IGM \citep{Basu04}, where $T_S$
is much smaller and varies with redshift similar to $T_{\rm CMB}$.

\begin{figure}[htb]
\includegraphics[scale = 0.43]{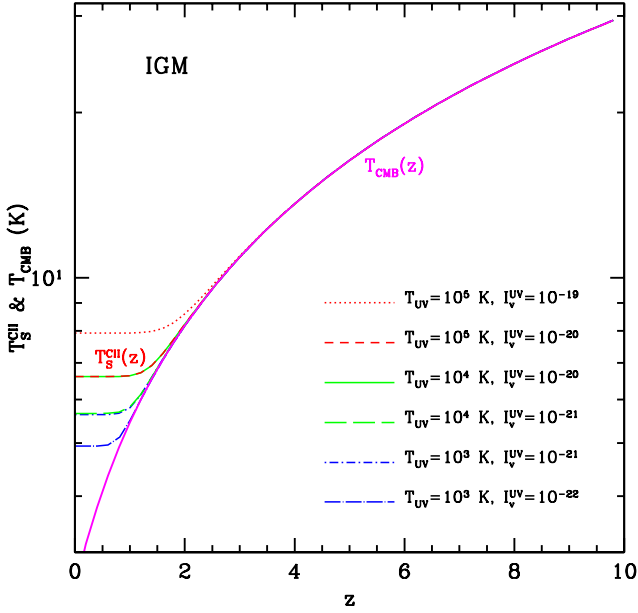} 
\caption{\label{fig:T_s_IGM} The CII spin temperature $T_S^{\rm CII}$ and 
$T_{\rm CMB}$ in the diffuse IGM as a function of the redshift. We
select several pairs of the UV color temperature
$T_{\rm UV}$ and the UV background intensity 
$I_{\nu}^{\rm UV}$ for this calculation.
We find the spin temperature is close to 
$T_{\rm CMB}$ at high redshifts ($z>2$).}
\end{figure}

\subsection{The CII spin temperature in the diffuse IGM}

In the IGM, the collisional process becomes unimportant since the number 
density of electrons and other elements are much smaller than in dense regions within the ISM of galaxies \citep{Basu04}. Also, the spontaneous emission and the stimulated 
absorption and emission by the CMB photons are considerable. We will also take into
account UV pumping that can enhance CII emission. This effect is similar to the Wouthuysen-Field effect for the 21-cm line \citep{Wouthuysen52,Field58,Hernandez06}
and, to the extent we are aware of, has not been discussed in the literature previously.

At high redshifts the soft UV background at $\rm 1330\AA$ generated by the first galaxies and
quasars can pump the CII ions from the energy level $2s^22p\ ^2P_{1/2}$ to 
$2s2p^2\ ^2D_{3/2}$ at $\rm 1334.53\AA$ and $2s^22p\ ^2P_{3/2}$ to $2s2p^2\ ^2D_{3/2}$ 
at $\rm 1335.66\AA$. Then this pumping effect can lead to the CII fine-structure
transitions $^2D_{3/2}\to ^2P_{3/2}\to ^2P_{1/2}$, which would mix the levels of the CII line at ${\rm 157.74 \mu m}$.
The UV de-excitation and excitation rates are given by \citep{Field58}
\be
P^{\rm UV}_{\rm ul}=\frac{g_{\rm k}}{g_{\rm u}}\frac{A_{\rm kl}}{\sum_{\rm n}A_{\rm kn}}A_{\rm ku}\bigg(\frac{c^2I_{\nu}^{\rm UV}}{2h\nu^3} \bigg)_{\nu_{\rm ku}}
\ee
and
\be
P^{\rm UV}_{\rm lu}=\frac{g_{\rm k}}{g_{\rm l}}\frac{A_{\rm ku}}{\sum_{\rm n}A_{\rm kn}}A_{\rm kl}\bigg(\frac{c^2I_{\nu}^{\rm UV}}{2h\nu^3} \bigg)_{\nu_{\rm kl}},
\ee
where ``k'' stands for the level $^2D_{3/2}$, $g_{\rm k}=4$, 
$A_{\rm kl}=2.41\times 10^8\ \rm s^{-1}$ and $A_{\rm ku}=4.76\times 10^7\ \rm s^{-1}$
are the Einstein coefficients of $^2D_{3/2}\to ^2P_{1/2}$ and $^2D_{3/2}\to ^2P_{3/2}$
respectively (see the NIST website 
\footnote{$\rm http://physics.nist.gov/PhysRefData/ASD/lines\_form.html$}),
and $\sum_{\rm n}A_{\rm kn}=A_{\rm kl}+A_{\rm ku}$.
Also, it is helpful to define the UV color temperature $T_{\rm UV,ul}$ in terms of 
the ratio of the UV de-excitation and excitation rates
\be
\frac{P_{\rm lu}^{\rm UV}}{P_{\rm ul}^{\rm UV}}=\frac{g_{\rm u}}{g_{\rm l}}{\rm exp}[-T_{\star,\rm ul}/T_{\rm UV,ul}].
\ee

Then, following a similar derivation as to the case of spin temperature in the ISM of galaxies,
we find that the CII spin temperature in the diffuse IGM is
\be
\frac{T_{\star,\rm ul}}{T_{S,\rm ul}} = log\Bigg\{ \frac{A_{\rm ul}[1+(I_{\nu}c^2/2h\nu^3)_{\nu_{\rm ul}}]+P_{\rm ul}^{\rm UV}}{A_{\rm ul}(I_{\nu}c^2/2h\nu^3)_{\nu_{\rm ul}}+P_{\rm ul}^{\rm UV}{\rm exp}(-T_{\star,\rm ul}/T_{\rm UV})} \Bigg\}. \label{eq:T_s_IGM}
\ee

In Fig.~\ref{fig:T_s_IGM}, we show the departure of $T_{S}$ from
$T_{\rm CMB}$ as a function of the redshift. We see that the
$T_{S}^{\rm IGM}$ does not change if $T_{\rm UV}>10^3\ \rm K$ when fixing
$I_{\nu}^{\rm UV}$, and it is close to $T_{\rm CMB}$ at high redshifts
($z>2$). Together with Eq.~\ref{eq:T_s_IGM} this implies that $A_{\rm ul}\gg P_{\rm ul}^{\rm UV}$, i.e. the spontaneous and stimulated emission are much 
greater than the UV pumping effect at high redshifts. By comparing 
Fig.~\ref{fig:T_s_IGM} with Fig.~\ref{fig:T_s_gal}, we can see
that the CII spin temperature in the ISM of galaxies is much larger than that of 
the diffuse IGM, and $T_{S}^{\rm IGM}$ is quite close to $T_{\rm CMB}$ at high
redshift while $T_S^{\rm gal}$ is larger than  $T_{\rm CMB}$.

\section{The calculation of the CII mean intensity}

To estbalish the overall intensity of the CII line emission
we calculate the distortion of the CMB spectrum $\Delta I_{\nu}$
and take into account the mean intensity of
the CII line emitted by the ISM galaxies and the IGM. Considering the 
spontaneous, stimulated emission and absorption, and
the expansion of the Universe, we write the radiative transfer equation as
\be \label{eq:dIv}
\frac{dI_{\nu}(z)}{ds}=j_{\nu}(z)-\alpha_{\nu}(z)I_{\nu}-3\frac{H(z)}{c}I_{\nu},
\ee
where $ds$ is the line element along the line of sight, and $H(z)$ is the
Hubble parameter. The spontaneous emission and absorption coefficients are 
$$j_{\nu}(z)=\frac{h\nu_{\rm ul}}{4\pi}n_{\rm u}(z)A_{\rm ul}\phi(\nu)$$
and
$$\alpha_{\nu}(z)=\frac{h\nu_{\rm ul}}{4\pi}\phi(\nu)(n_{\rm l}B_{\rm lu}-n_{\rm u}B_{\rm ul})$$
respectively, where $\phi(\nu)$ is the line profile function which can be
set as delta function $\phi(\nu')=\delta(\nu'-\nu)$ if the thermal broadening
or velocity is much smaller than the frequency resolution 
\citep{Suginohara99,Basu04}. 
Integrating Eq.\ref{eq:dIv} along the line of sight then gives
\ba
\Delta I_{\nu} &=& \int \frac{j_{\nu}(z)-\alpha_{\nu}(z)I_{\nu}}{(1+z)^3}ds \\ \nonumber
               &=& \int \frac{j_{\nu}(z)-\alpha_{\nu}(z)I_{\nu}}{H(z)(1+z)^4}cdz.
\ea

Using the relation of the line profile and the redshift
$\phi(\nu')=\phi[\nu_0(1+z')]=[(1+z)/{\nu}]\delta(z'-z)$
we can obtain the
integrated intensity of the CII emission lines at z as
\ba
\Delta I_{\nu} = \frac{hc}{4\pi H(z)(1+z)^3}A_{\rm ul}\ f_{\rm CII}^{\rm grd}n_{\rm CII}(z)\times \\ \nonumber
\frac{g_{\rm u}}{g_{\rm l}}{\rm exp}(-T_{\star,\rm ul}/T_{S,\rm ul})\bigg[ 1-\frac{{\rm exp}(T_{\star,\rm ul}/T_{S,\rm ul})-1}{(2h\nu^3/c^2I_{\nu})_{\nu_{\rm ul}}} \bigg],
\ea
where $f_{\rm CII}^{\rm grd}$ is the fraction of CII ions at the ground
level $^2P_{1/2}$. If $T_S\gg T_{\star}(>T_{\rm CMB})$, which is the usual case in the galaxies, then
${\rm exp}(\pm T_{\star,\rm ul}/T_{S,\rm ul})\sim 1$, and we can
find
\be
\Delta I_{\nu} = \frac{hc}{4\pi H(z)(1+z)^3}\frac{g_{\rm u}}{g_{\rm l}}A_{\rm ul}\ f_{\rm CII}^{\rm grd}n_{\rm CII}(z).
\label{eq:Iv_simp}
\ee 

If $T_{S}$ is much larger than 
$T_{\star, \rm ul}$ (e.g. the cases in galaxies), $f_{\rm CII}^{\rm grd} \simeq 1/3$
is a good approximation, and when $T_{S}\ll T_{\star, \rm ul}$ we can
set $f_{\rm CII}^{\rm grd} \simeq 1$ (e.g. the case of the IGM). 
The $n_{\rm CII}(z)$ is the number density of the CII ions at $z$, which has to be carefully estimated for both the galaxy and IGM cases.

To calculate the CII intensity, we now need an estimate of the
number density of the CII ions. For the IGM case, it is easy to
calculate if we know the  metallicity evolution and the average 
baryonic density as a function of z. However, for the ISM of galaxies, we also have
to find the fraction of the CII ions which exceed the critical
density to trigger the dominant collisional emission.
In the next section, we will first evaluate the $n_{\rm CII}$ theoretically
for both the ISM and IGM cases, and then calculate $n_{\rm CII}$ using 
data from a simulation and compare it to the analytical result as a check on the
consistency. We also provide an order of magnitude estimate on the $z \sim 6$ to 8 CII mean intensity using
scaling relations such as those involving star-formation rate and the total far-IR luminosity of galaxies.

\subsection{The analytic estimation}

We start by writing
\be
n_{\rm CII}(z)=f_{\rm CII}(z)Z_{\rm C}\bar{n}_{\rm gas}(z),
\label{eq:n_CII}
\ee
where $Z_{\rm C}=X_{\rm C}Z_{\odot}^{\rm C}$, 
$Z_{\odot}^{\rm C} = 3.7\times 10^{-4}$ is the solar carbon 
abundance \citep{Basu04}, and we assume $X_{\rm C}=1$ in the galaxy
and $X_{\rm C}=10^{-2}$ in the IGM \citep{Savaglio97,Aguirre05,Kramer10}.
The $f_{\rm CII}(z)=Z/Z_{\odot}(z)$ is the global metallicity 
at different redshift, and we use an approximated relation 
$Z/Z_{\odot}(z)=10^{-0.4z}$ in our calculation, which assumes
the present mean metallicity of the Universe is equal to the solar 
metallicity (note that this assumption may overestimate the metallicity
at low redshifts) and the carbon atoms are totally ionized to be CII. 
This relation is consistent with the observational data
from the damped Ly-$\alpha$ absorbers (DLAs) metallicity measurements,
which covers the redshift range from 0.09 to 3.9 \citep{Kulkarni05}, 
and also matches a previous theoretical estimation \citep{Pei95,Pei99,Malaney96}. 

The $\bar{n}_{\rm gas}(z)$ in Eq.~\ref{eq:n_CII} is the mean number 
density of the gas, which, for the IGM case, is just the baryon density, 
$\bar{n}_{\rm b}(z) = 2\times 10^{-7} (1+z)^3 {\rm cm^{-3}}$.
For the ISM of galaxies, we use 
$\bar{n}_{\rm gas}(z)=f_{\rm gas}^{\rm hot}f_{\rm gas}^{\rm cr}\bar{n}_b(z)$, 
where $f_{\rm gas}^{\rm cr}$ is the fraction of the gas that are 
present in dense environments of the ISM and satisfies 
$n_{\rm gas}\gtrsim n_{\rm cr}$ \citep{Suginohara99}.
The value of $f_{\rm gas}^{\rm cr}$ depends on the gas clumping within 
the ISM of galaxies and the Jeans mass \citep{Fukugita94,Suginohara99}. 
We find $f_{\rm gas}^{\rm cr}$ is about 0.1 at $z=6$ and decreases slowly 
at higher redshifts in our simulation \citep{Santos10}.
For simplicity, we just take $f_{\rm gas}^{\rm cr}=0.1$ to be the case for all 
galaxies independent of the redshift. As is clear, this parameter is the 
least uncertain of the calculation we present here.
Observations with ALMA and other facilities of {\it Herschel} 
galaxy samples will eventually narrow down this value.

\begin{figure}[htb]
\includegraphics[scale = 0.43]{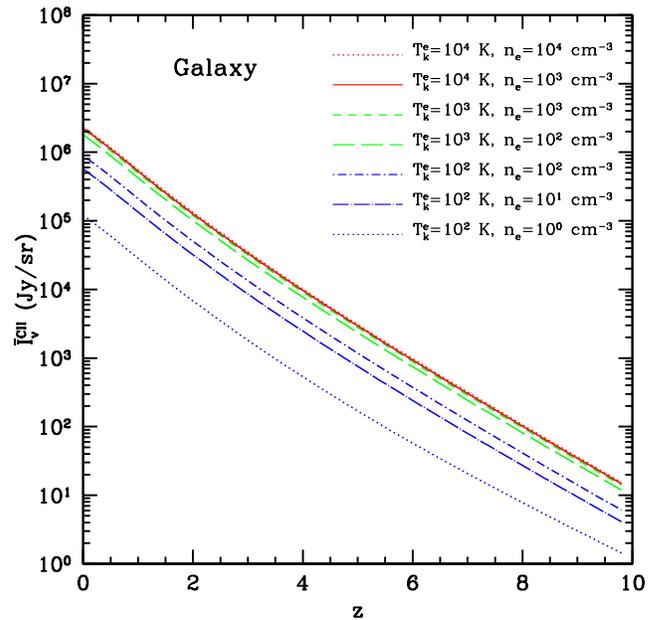} 
\caption{\label{fig:Iv_gal} The mean intensity of CII
emission line from the ISM gas in galaxies as a function of the redshift.
We find that the electron collisional emission
saturates when $n_{\rm e}\gtrsim 10^2\ \rm cm^{-3}$
and $T^{\rm e}_{\rm k}\gtrsim 10^3\ \rm K$. 
}
\end{figure}

The fraction of the ``hot'' gas ($T>10^3$ K) in halos 
$f_{\rm gas}^{\rm hot}$ is also included here, since the main contribution of CII emission 
comes from the gas with $T>10^3$ K (see Fig.~\ref{fig:Iv_gal}). 
We find the fraction is around 0.3 and remains constant from $z=8$ to 6 
in the De Lucia catalog \citep{DeLucia07},
which is used to derive the CII number density from the simulation
in the next subsection.
As we are computing the mean intensity expected in a cosmological survey,
$\bar{n}_{\rm gas}$ is the number density for galaxies in a large space 
volume instead of one individual galaxy and thus we can still use
$\bar{n}_{\rm b}$ for the ISM of galaxies.

\begin{figure}[!b]
\includegraphics[scale = 0.43]{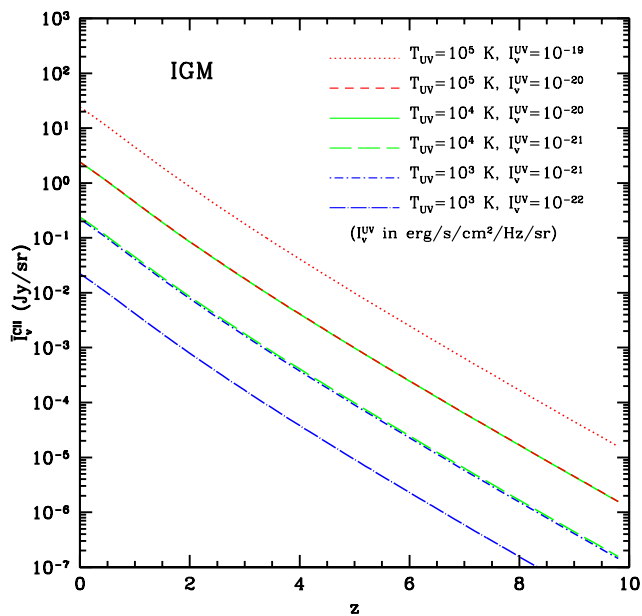} 
\caption{\label{fig:Iv_IGM} The mean intensity of CII
emission line from carbon in the IGM as a function of the redshift.
The intensity here is much smaller than that of ISM
in galaxies since the spin temperature is close to the 
CMB temperature and the gas density is much lower than galaxy ISM.}
\end{figure}

In Fig.~\ref{fig:Iv_gal}, we plot an analytical estimate of intensity of CII emission 
line from the ISM of galaxies as a function of redshift. We select the same pairs 
of $T_{\rm k}^{\rm e}$ and $n_{\rm e}$ as in the Fig.~\ref{fig:T_s_gal}.
We see that the intensity is practically independent of the electron 
number density and the temperature when $n_{\rm e}>10^2\ \rm cm^{-3}$ and 
$T^{\rm e}_{\rm k}>10^3\ \rm K$, e.g., the signal is only seen in emission 
and is essentially proportional to the CII density (see Eq.~\ref{eq:Iv_simp}).
Even for $n_{\rm e}=1$ and $10\ \rm cm^{-3}$ which is less than the 
$n_{\rm e}^{\rm cr}$, we still find a significant intensity for the CII emission from the ISM of galaxies.

In Fig.~\ref{fig:Iv_IGM}, we show the same for carbon in the diffuse IGM.
We find the $I_{\nu}^{\rm CII}$ from the IGM is much smaller than
that from galaxies, and  $I_{\rm CII}^{\rm gal}/I_{\rm CII}^{\rm IGM}\gtrsim 10^4$ for all 
cases we consider at all redshifts. 
This is because the CII spin temperature and the CII abundance in the ISM of galaxies 
are much larger than that in the diffuse IGM.
Thus the CII emission from IGM can be safely neglected, and hereafter we 
will just take into account the CII emission from the ISM of galaxies when discussing intensity fluctuations.

Note the line intensity measurements of individual galaxies are generally described with PDR models
using the number density and the UV intensity within the ISM, instead of number density and temperature as we use here \citep{Malhotra01}.
We depart from the PDR modeling approach as we are
considering global models that are appropriate for cosmological observations and are not attempting to model the properties of line emission from individual
galaxies. It is likely that future work will need to improve our model by making use of a more accurate description of the stellar mass function and the
star-formation history of each of the galaxies in a large cosmological simulation by performing calculations to determine $T_{\rm e}$ given the
UV intensity field dominated by the massive, bright stars.
When making predictions related to the CII intensity power spectrum (Section~4), we  take as default values
$T_{\rm e}=1000$K and $n_{\rm e}=100$ cm$^{-3}$. These values are fully consistent with the findings of \citet{Malhotra01}, where they
find for 60 normal star-forming galaxies, the values of $T \sim 225$ to 1400K and $n \sim 10^2$ to 10$^4$ cm$^{-3}$.

\subsection{Intensity estimated from numerical simulations}

Since the evolution of the metallicity and 
the critical fraction of the gas $f_{\rm gas}^{\rm cr}$ are hard to estimate, 
the number density of the CII ions $n_{\rm CII}$ is not well determined
analytically for the case involving CII emission from the ISM of galaxies.
So we use results derived from a numerical
simulation to check the analytical estimation. To do this we  first derive the expected CII mass, 
$M_{\rm CII}$, in a halo with mass $M$ at a given redshift. Then, by integrating over all possible halo masses in a given volume, we estimate the CII mass for that same volume.

\begin{figure}[htb]
\centerline{
\includegraphics[scale = 0.35]{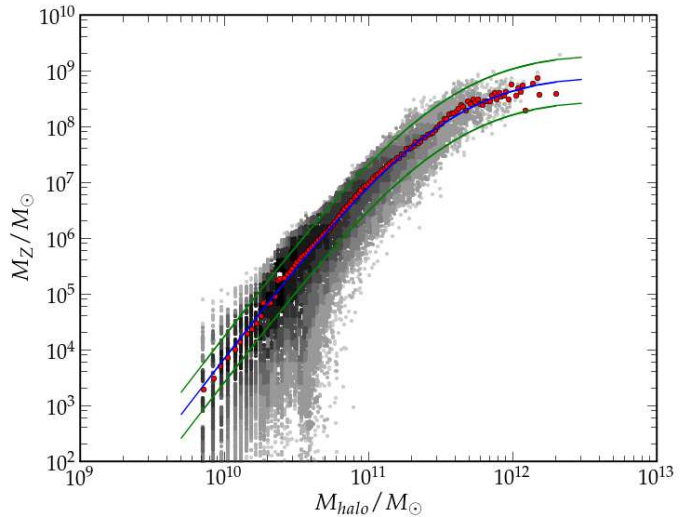} 
}
\caption{ The mass in metals $M_Z$ as a function of the 
halo mass $M$ at $z=8$. The solid lines shows the mean 
relation (blue center line) and $\pm 1\sigma$ relation 
(green lines). The red dots denote the mean value of 
the scattering (gray region) when binned to 150 
logarithmic intervals in halo mass.}
\label{fig:metal_hot_gas}  
\end{figure}

To find  $M_{\rm CII}(z)$ with simulations, we make use of the galaxy catalog 
from \cite{DeLucia07}. This catalog is obtained by post-processing the Millennium 
dark matter simulation with semi-analytical models of galaxy formation \citep{Springel05},
and has the metal content in each of four different components: stars, cold gas, 
hot gas and an ejected component. In this calculation we will assume that the CII 
emission comes from the hot gas component which has a temperature of about 
$10^5$ - $10^7$ K. However, according to \cite{Obreschkow09a}, some of the gas 
in galactic disks which is considered as cold gas 
(T $\approx$ $10^2 - 10^3\ \rm K$) in the De Lucia catalog, should actually be warm (T $\approx$ $10^4\ \rm K$) 
and ionized.   So we will also consider the case where $10\%$ of this cold gas from the \cite{DeLucia07} simulation is recategorized as warm and thus contribute
to the CII luminosity. 

In Fig.~\ref{fig:metal_hot_gas} we plot the mass in metals 
$M_Z$ as a function of the halo mass $M$ for the hot gas only case. 
There is some scatter, specially at the low mass end, but note that 
if the experimental volume is large enough, we will be observing a 
large number density of galaxies such 
that the total $M_Z$ should approach the average. 
The average relation between $M_Z$ and halo mass $M$ can 
be parameterized in the form $M_Z(M)=M_0(M/M_c)^b(1+M/M_c)^{-d}$. At $z=6$, $z=7$ 
and$z=8$, these parameters take the values $M_0=1.1\times10^9$, 
$1.0\times10^9$ and $1.6\times10^9$, 
$M_c=3.5\times10^{11}$, $3.5\times10^{11}$ and $3.7\times10^{11}$, $b=3.6$, $3.4$ 
and $3.4$, and $d=3.25$, $3.1$ and $3.6$, respectively.

At high ISM temperatures with $T > 100$K we assume carbon is ionized, so we 
have $M_{\rm CII}$=$f^{C}_{\odot}M_Z$ where $f^{C}_{\odot}=0.21$ is the carbon 
fraction of the mass in metals in the Sun \citep{Arnett96}. 
By taking into account the expected number of halos,
we can obtain $n_{\rm CII}$ as
\be
n_{\rm CII}^{\rm sim}(z)=\int^{M_{\rm max}} _{M_{\rm min}} dM \frac{dn}{dM} \frac{M_{\rm CII}(M,z)}{m_{\rm c}},
\label{eq:2}
\ee
where $m_c$ is the atomic carbon mass, $M$ is the halo mass and $dn/dM$ is 
the halo mass function (Sheth $\&$ Tormen 1999). The integration is made 
from $M_{\rm min}=10^8\ \rm M_{\odot}/h$ to $M_{\rm max}=10^{13}\ \rm M_{\odot}/h$,
and $M_{\rm min}$ is the minimum mass of the dark matter halo that can host 
galaxies. Note that the result is insensitive to the exact value of $M_{\rm max}$ since 
halos with $M>10^{13}$ are rare due to the exponential term in the halo mass function.
From the previous section we can safely assume that the spin temperature is saturated (e.g. the CII is only seen in emission) so that we now
have all the ingredients to calculate the signal using the simulation values.

\begin{figure}[htb]
\centerline{
\includegraphics[scale = 0.45]{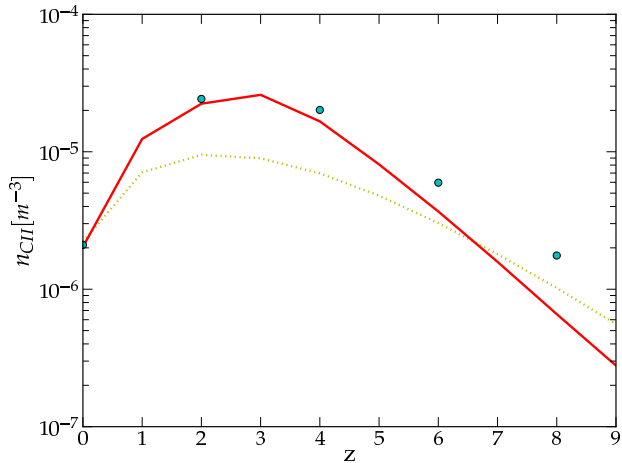} 
}
\caption{The CII proper number density as a function of 
redshift. The red solid line is derived from our simulation 
with just hot gas contributing to $L_{\rm CII}$, 
the yellow dotted line is our analytic result, and the blue 
spots are obtained from the simulation using both the hot gas and the warm gas ($10\%$ of the cold gas in the galaxies; see text for details).}
\label{fig:n_CII}  
\end{figure}

\begin{figure}[htb]
\centerline{
\includegraphics[scale = 0.48]{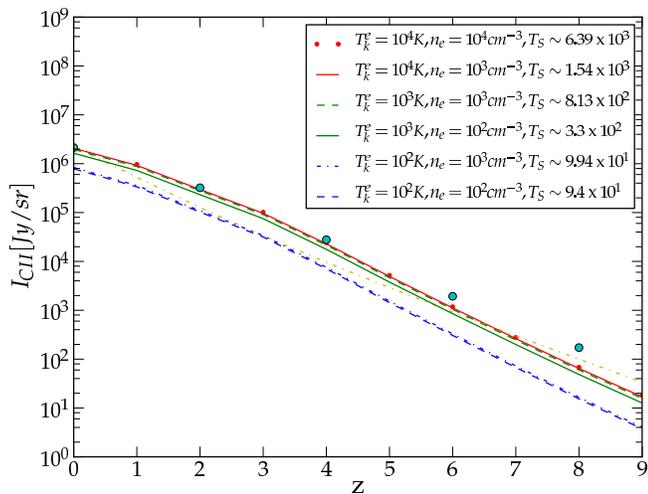} 
}
\caption{The mean intensity of CII emission line from the outputs 
obtained from the \cite{DeLucia07}
simulation at several different redshifts. The yellow dotted line 
is our analytic result, and the blue spots are obtained from 
the simulation with the hot and warm gas 
contributing to the $L_{\rm CII}$.
The yellow line and the blue spot were calculated with
$n_{\rm e}=10^3\ \rm cm^{-3}$ and 
$T_{\rm k}^{\rm e}=10^4\ \rm K$. The other lines are derived 
from the simulation assuming only the hot gas contributes to 
the $L_{\rm CII}$.}
\label{fig:ICII}
\end{figure}

In Fig.~\ref{fig:n_CII}, we show the number density of the CII ions 
$n_{\rm CII}^{\rm sim}(z)$ using the simulation result. We find our analytic 
result (black dashed line) is well consistent with that derived from the simulation especially 
at high redshift. The departure at low redshift in our analytical calculation is reasonable since the
the hot gas fraction $f_{\rm gas}^{\rm hot}$ should be higher at low
redshifts because of the higher star formation rate.

Assuming only the hot gas component of \cite{DeLucia07}  contribute to the CII line luminosity, $L_{\rm CII}$, 
 our two calculational methods differ by a factor of a few; however, after adding the 
warm component, the analytical result of the previous section match the numerical values
at $z > 6$ where we are especially interested. The difference is again same for the
mean intensity of the CII emission shown in Fig.~\ref{fig:ICII}. Again, we find that the 
CII emission is saturates when $n_{\rm e}\gtrsim$ $10^2\ \rm cm^{-3}$ 
and $T^{\rm e}_{\rm k}\gtrsim 10^3\ \rm K$. For higher values, CII intensity is almost independent of $n_{\rm e}$ and 
$T^{\rm e}_{\rm k}$.

\subsection{Intensity estimated from observed scaling relations}

At $z \sim 7$, our calculations involving either analytical or through \cite{DeLucia07} simulations suggest a mean intensity of about  100 to 500 Jy/sr with a preferred
value around 200 Jy/sr. We can use an approach based on observed scaling relations, similar to the approach used by \cite{Carilli11} to estimate the mean CO(1-0) intensity
during reionization, to estimate the mean CII intensity. We begin with the star-formation rate from $z \sim 6$ to 8. While estimates exist in the literature from Lyman break galaxy (LBG)
dropouts \citep{Bouwens08}, such estimates only allow a lower estimate of the star-formation rate as the luminosity functions are limited to the bright-end of galaxies and the slope at the
low-end is largely unknown and could be steeper than indicated by existing measurements. An independent estimate of the star-formation rate during reionization involve the use of
gamma-ray bursts \citep{Kistler09}. Together, LBG and GBR star-formation rates lead to a range of  0.01 to 0.1 M$_{\sun}$ yr$^{-1}$ Mpc$^{-3}$ at $z \sim 7$. 

We can convert this SFR to the
CII luminosity of $\sim 3\times 10^{40}$ to $\sim 3\times 10^{41}$  ergs s$^{-1}$ using the observed scaling relation of \cite{DeLooze11} for low-redshift galaxies
when averaging over 100 Mpc$^3$ volumes. We also
get a result consistent with this estimate when converting the SFR to total integrated FIR luminosity and then assuming that 10$^{-2}$ to $10^{-3}$ of the L$_{\rm FIR}$ appears in CII \cite{Stacey10}, 
again consistent with $z\sim 2$ redshift galaxy samples. Once again we note that our estimate is uncertain if either the SFR to CII luminosity calibration evolves with redshift
or the CII to FIR luminosity ratio evolves with redshift. The above range with an order of magnitude uncertainty in the SFR, and subject to uncertain evolution in observed
scaling relations from $z \sim 2$ to 7, corresponds to an intensity of 40 to 400 Jy/sr at $z \sim 7$. This range is consistent with our independent estimate, but could be improved in the near future
with CII and continuum measurements of high redshift galaxy samples with ALMA.

\section{The CII intensity power spectrum}

The CII intensity we have calculated is just the mean intensity, so in this Section we will discuss spatial variations in the intensity and consider 
the power spectrum of the CII emission line as a measure of it. This power spectrum captures the underlying matter distribution and if CII line intensity fluctuations
can be mapped at $z > 6$, the line intensity power spectrum can be used to probe the spatial distribution of galaxies present during reionization.

Since the CII emission from the ISM of galaxies will naturally trace the underlying cosmic matter density field, we can write the
CII line intensity fluctuations due to galaxy clustering as
\be
\delta I_{\nu}^{\rm CII}=\bar{b}_{\rm CII}\bar{I}_{\nu}^{\rm CII}
\delta({\rm \bf x}),\label{eq:d_Iv}
\ee
where $\bar{I}_{\nu}^{\rm CII}$ is the mean
intensity of the CII emission line from the last section, 
$\delta({\rm \bf x})$ is the matter over-density at the location
$\rm {\bf x}$, and $\bar{b}_{\rm CII}$ is the average galaxy bias 
weighted by CII luminosity (see e.g. \citealt{Gong11}).

Following Eq.~\ref{eq:2} and taking into account that the fluctuations in the halo number density will be a biased tracer of the dark matter, 
the average bias can be written as \citep{Visbal10}

\be
\bar{b}_{\rm CII}(z)=\frac{\int^{M_{\rm max}}_{M_{\rm min}} dM \frac{dn}{dM} M_{\rm CII}(M) b(z,M)}{\int^{M_{\rm max}}_{M_{\rm min}} dM \frac{dn}{dM} M_{\rm CII}(M)},
\label{eq:4}
\ee 
where $b(z,M)$ is the halo bias and $dn/dM$ is the
halo mass function \citep{Sheth99}. We take $M_{\rm min}=10^8\ \rm M_{\odot}/h$
and $M_{\rm max}=10^{13}\ \rm M_{\odot}/h$.
Then we can obtain the clustering power spectrum of the CII emission line
\be
P_{\rm CII}^{\rm clus}(z,k) = \bar{b}_{\rm CII}^2 \bar{I}_{\rm CII}^2 P_{\delta \delta}(z,k),
\ee
where $P_{\delta \delta}(z,k)$ is the matter power spectrum. 

The shot-noise power spectrum, due to discretization of the galaxies,
is also considered here. It can be written as \citep{Gong11}
\be
P^{\rm shot}_{\rm CII}(z) = \int_{M_{\rm min}}^{M_{\rm max}} dM \frac{dn}{dM} 
\bigg(\frac{L_{\rm CII}}{4\pi D_{\rm L}^2}y(z)D_{\rm A}^2\bigg)^2,
\ee
where $D_{\rm L}$ is the luminosity distance, $D_{\rm A}$ is the comoving
angular diameter distance and $y(z)={d\chi}/{d\nu}={\lambda_{\rm CII}(1+z)^2}/{H(z)}$, 
where $\chi$ is the comoving distance, $\nu$ is the observed
frequency, $\lambda_{\rm CII}$ is the rest frame wavelength of the CII line.
The $L_{\rm CII}$ is the CII luminosity which can be derived from 
the $I_{\nu}^{\rm CII}$, and we assume
$L_{\rm CII}(M,z)=B\times M_{Z}(M,z)\ [L_{\odot}]$
and finally find $B=100.63$, $100.36$, and $100.17$ at $z=6$, $7$, and $8$
respectively. The total CII power spectrum is
$P_{\rm CII}^{\rm tot}(z,k)=P_{\rm CII}^{\rm clus}(z,k)+P_{\rm CII}^{\rm shot}(z)$.

\begin{figure*}[htb]
\epsscale{1.9}
\centerline{
\resizebox{!}{!}{\includegraphics[scale = 0.4]{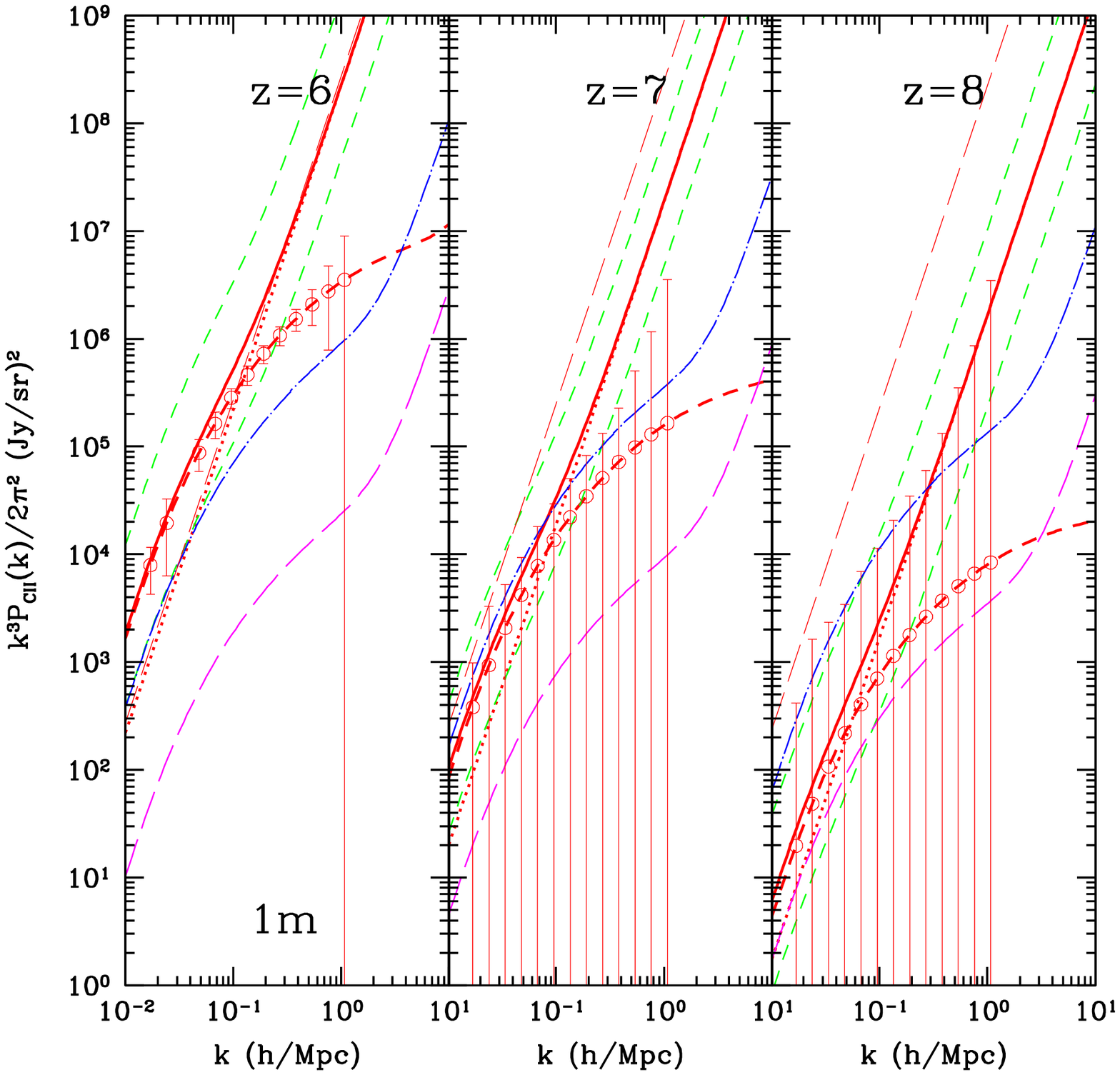}}
\resizebox{!}{!}{\includegraphics[scale = 0.4]{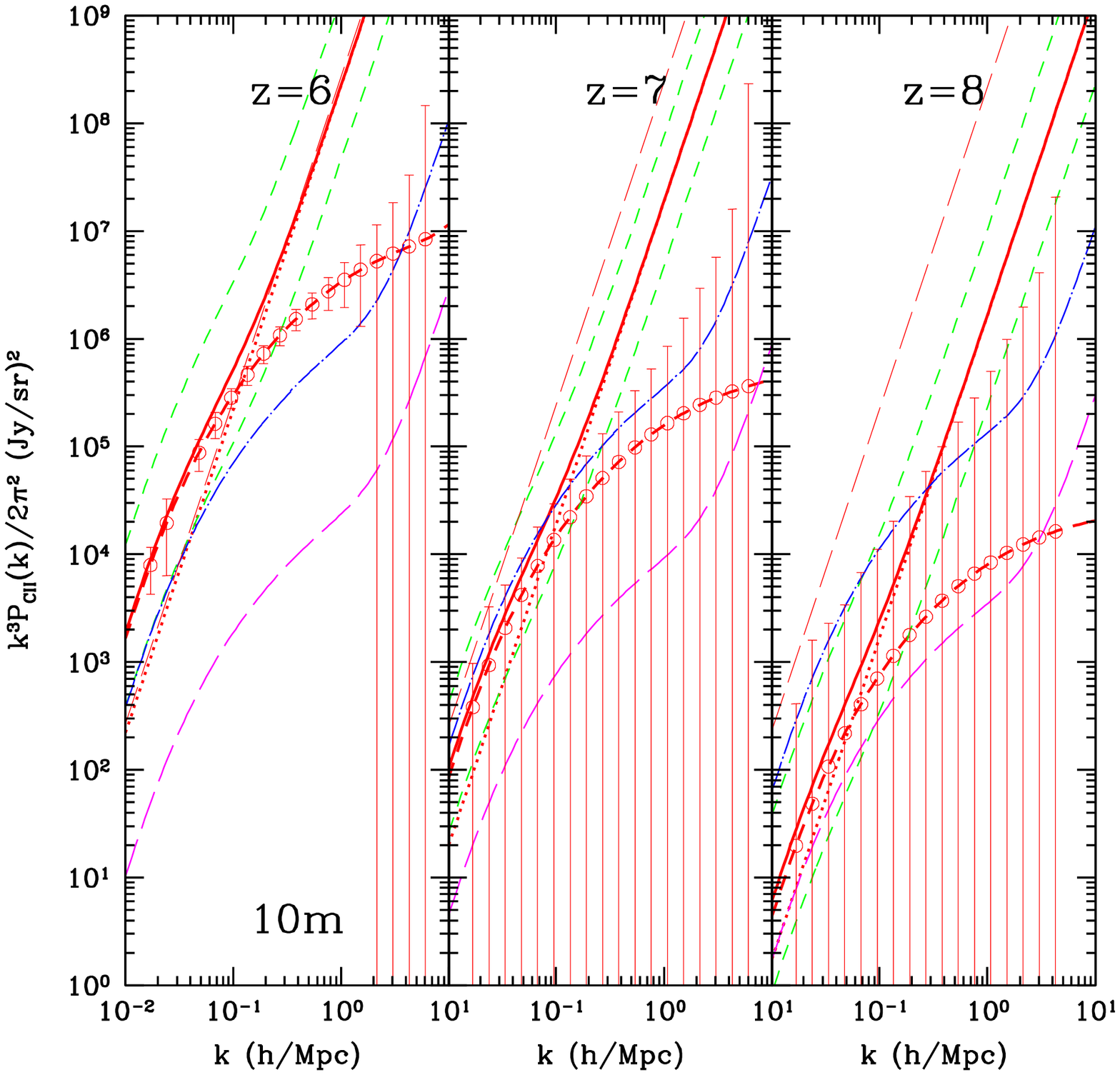}}
}
\centerline{
\resizebox{!}{!}{\includegraphics[scale = 0.4]{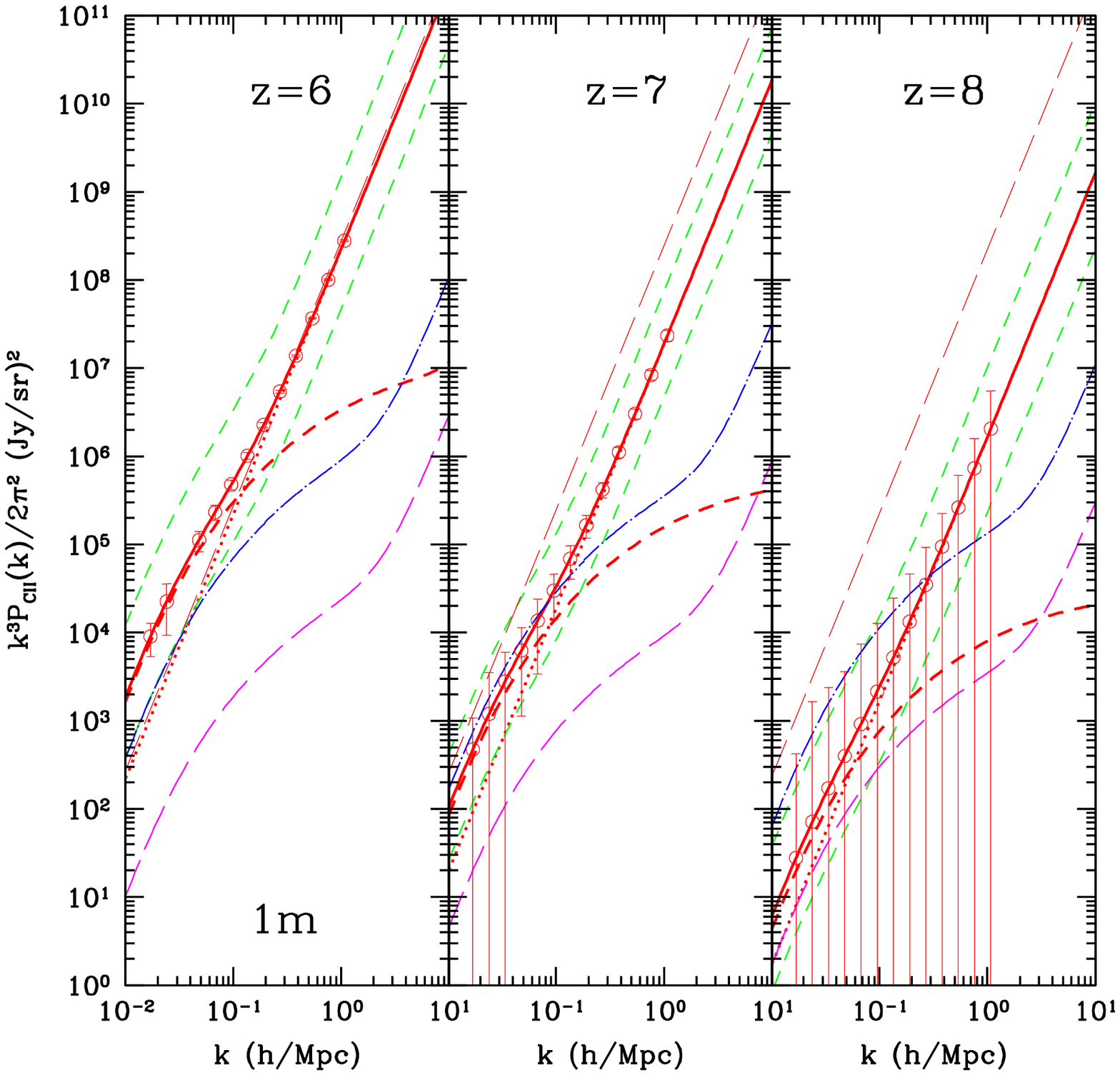}}
\resizebox{!}{!}{\includegraphics[scale = 0.4]{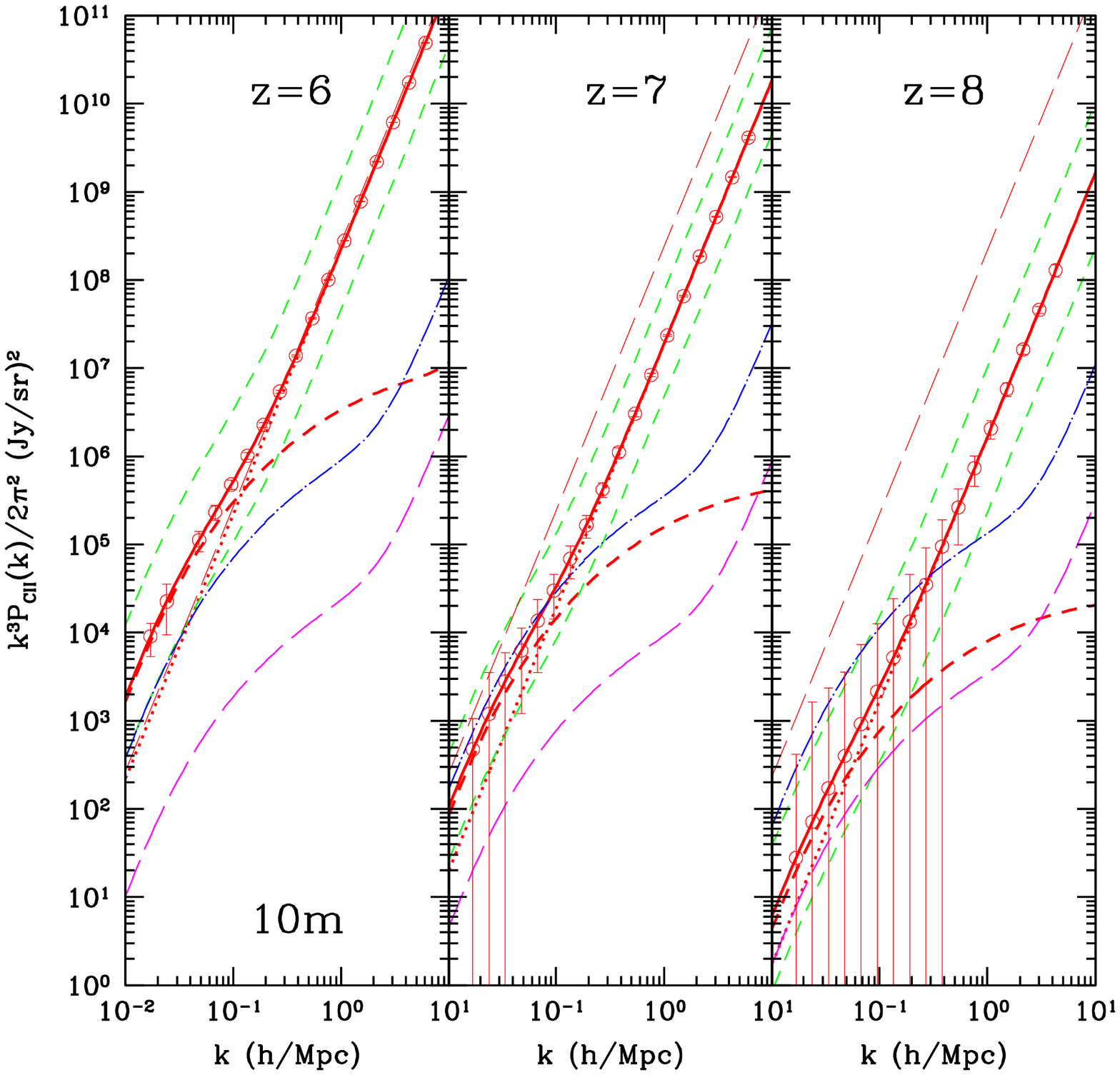}}
}
\epsscale{1.0}
\caption{\label{fig:P_CII} The clustering, shot-noise and
total power spectrum of CII emission line at $z=6$, $z=7$ and $z=8$. 
The red solid, dashed and dotted lines denote the CII total, clustering
and shot-noise power spectrum respectively. The $P_{\rm CII}^{\rm clus}$ 
is estimated 
from the simulation with only the hot gas contributing to $L_{\rm CII}$,
and we assuming that 
$T_{\rm k}^{\rm e}=10^3\ \rm K$ and $n_{\rm e}=10^2\ \rm cm^{-3}$ here. 
The green dashed lines are the $1 \sigma$ error of the CII power spectrum
which are derived from the $1 \sigma$ error of $M_z-M$ relation 
in Fig.~\ref{fig:metal_hot_gas}. 
The error bars and noise power spectrum (red long-dashed line) 
in the top left and top right panels are estimated with 1m and 
10m aperture for CII line respectively. The magenta long dashed
line is derived from the CII luminosity estimated by \cite{Visbal10}.
The blue dash-dotted line is estimated via the relation
$L_{\rm CII}/L_{\rm CO(1-0)}\simeq 10^4$. The bottom left and
bottom right panels are the same plots but with the error bars on the total
power spectrum.
}
\end{figure*}

In Fig.~\ref{fig:P_CII}, as an example we plot the 
clustering, shot-noise and total power spectrum of the CII emission
at $z=6$, $z=7$ and $z=8$. Using the $\tt Halofit$ code \citep{Smith03}, 
we calculate the non-linear matter power spectrum. 
In order to calculate the intensity of the CII line we use 
$T_{\rm k}^{\rm e}=10^3\ \rm K$ and $n_{\rm e}=10^2\ \rm cm^{-3}$, 
which are a possible representation of the conditions at which CII emits 
in the galaxies \citep{Malhotra01}. 

For comparison, the CII power spectrum estimated by
other works are also shown here. The blue dash-dotted line denotes the
$P_{\rm CII}^{\rm tot}$ derived from the relation 
$L_{\rm CII}/L_{\rm CO(1-0)}\simeq 10^4$ \citep{Breuck11}, and the 
$L_{\rm CO(1-0)}$ is from the calculation of \cite{Visbal10}.
The magenta long dashed line is the $P_{\rm CII}^{\rm tot}$ evaluated via
the CII luminosity derived from \cite{Visbal10}. 

As expected the CII power spectrum is larger than the CO(1-0) power spectrum calculated in \cite{Gong11}.
Note that the CO intensities predicted by \cite{Gong11} should be corrected by a factor of $1/(1+z)$ which comes from a missing conversion factor between the CO flux obtained from \cite{Obreschkow09b} and the actual CO(1-0) luminosity. Although this correction further reduces the CO signal, we point out that the model we assumed for the CO luminosity as a function of the halo mass generates a steep dependence 
at the low mass end. If we use a less steep model at the low end of halo masses ($M < 10^{10}\ \rm M_{\odot}/h$), where the simulation has a large scatter,
such as $L_{\rm CO(1-0)}\propto M_{\rm halo}$, then the CO signal can increase by a factor of a few, partially compensating the $1/(1+z)$ correction.

Comparing the CO(1-0) power spectrum (using the $1/(1+z)$ correction but keeping the luminosity model used in \cite{Gong11}) with the CII one
with its density estimated using only the hot gas component from the simulation, implies a luminosity relation of $L_{\rm CII}/L_{\rm CO(1-0)}\approx 4\times10^4$ 
when $z \sim 6$ reducing to below $10^4$ when $z \sim 8$.   The observational value for this relation as valid for $0 < z < 2$ is $L_{\rm CII}/L_{\rm CO(1-0)}=4100$ as obtained by 
\cite{Stacey91} (see also \citealt{Stacey10}).
However detections of CII emission from star-forming galaxies at $z > 6$ is almost non-existent and there is a strong possibility for an evolution with redshift for this relation.
For the handful of galaxies studied, the possibility for evolution is supported by the $L_{\rm CII}/L_{\rm CO(1-0)}\simeq 10^4$ value derived for
LESS J033229.4-275619 at $z=4.76$ \citep{Breuck11}, which is one of the highest redshift CII detections in a sub-mm selected, star-formation dominated galaxy by far, though
higher redshift detections are soon expected as part of {\it Herschel} followup campaigns.
In Fig.~\ref{fig:P_CII} we show that this observational ratio (blue dash-dotted line) is consistent with our direct CII prediction from the simulation with the hot
gas case (red solid line) around $1\sigma$ level (green dashed lines, calculated using the $1\sigma$ error of the $M_Z$-$M$ relation).
The noise power spectrum and the error bars are estimated for an assumed (sub-)milimeter survey, which we will discuss in some detail in Section 7. 

The luminosity formula in \cite{Visbal10} based on the observed sub-mm and mm emission line ratios (see below for CO version) can also be used to calculate the
the CII power spectrum as a function of the redshift, but it leads to a result (long dashed magenta line)
that is smaller than that estimated by the $L_{\rm CII}$-$L_{\rm CO(1-0)}$ relation (blue dash-dotted line) and our CII power spectrum. This is effectively due to a difference
in the calibration with CO(1-0) luminosity from M82 and CII luminosity from a sample of low-redshift galaxies \citep{Malhotra01}. 
The ratio $L_{\rm CII}$/$L_{\rm CO(1-0)}$ can be as low as 1.6$\times 10^3$ for luminous low-redshift galaxies such ULIRGs\cite{Stacey10}.
This sets a lower bound on the $L_{\rm CII}$ estimation if we assume that $z > 6$ galaxies are analogous to local ULIRGs.

In Fig.~9, we plot the intensity maps of the CII emission at $z=6$, 7 and 8 using  \cite{DeLucia07} simulations to calculate the CII line intensities
to show what the sky looks like at each of these redshifts when mapped with CII. The emission can be traced to individual galaxies. Each of these maps span 3 degrees in each direction
and the color bar to the right show the intensity scaling in units of Jy/sr. Note the large reduction in the maximum intensity from $z=6$ to $z=8$ due to the decrease in the overall abundance of
metal content in the ISM of galaxies at high redshifts.

\begin{figure}[htb]
\includegraphics[scale=0.34]{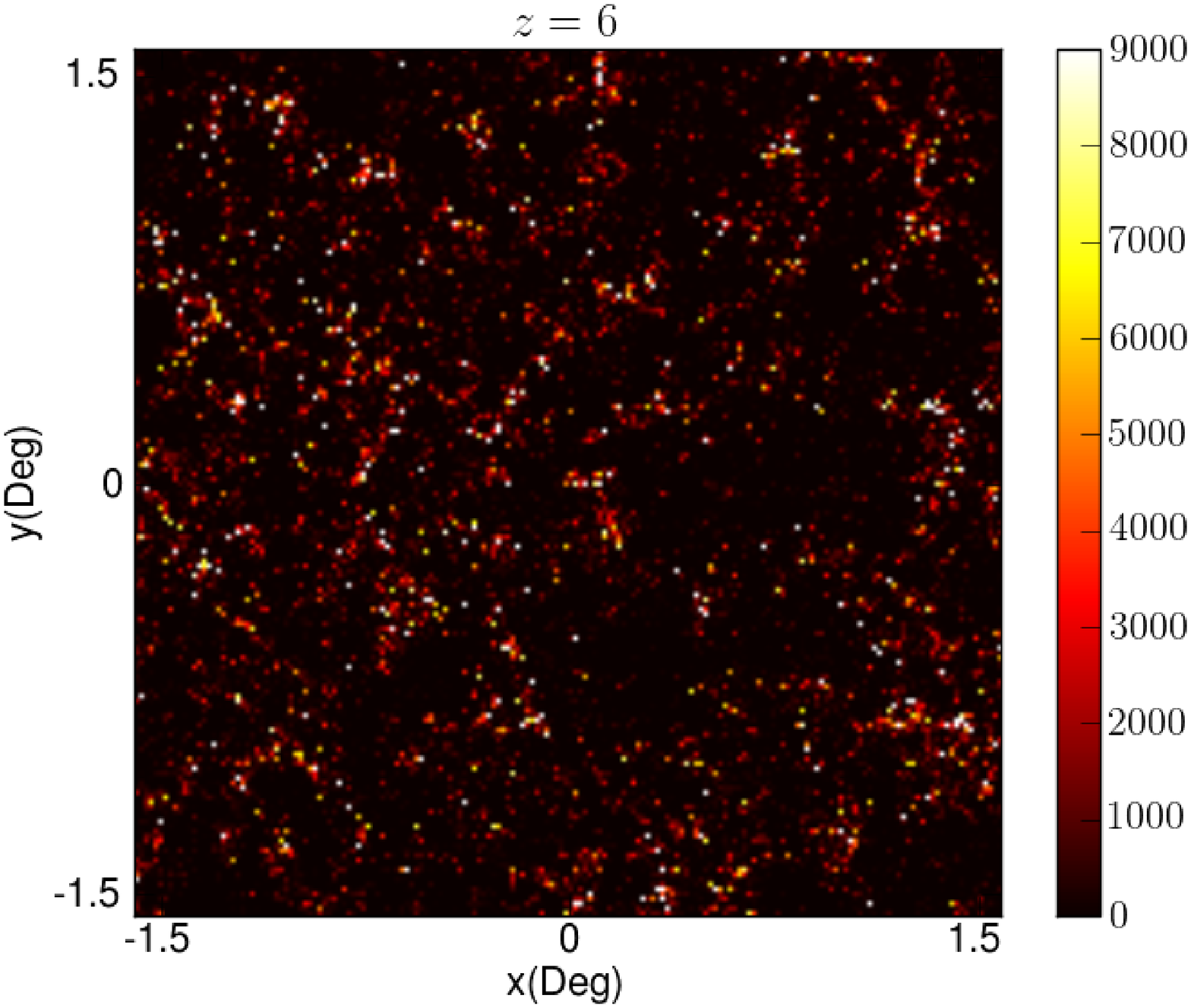}
\includegraphics[scale=0.34]{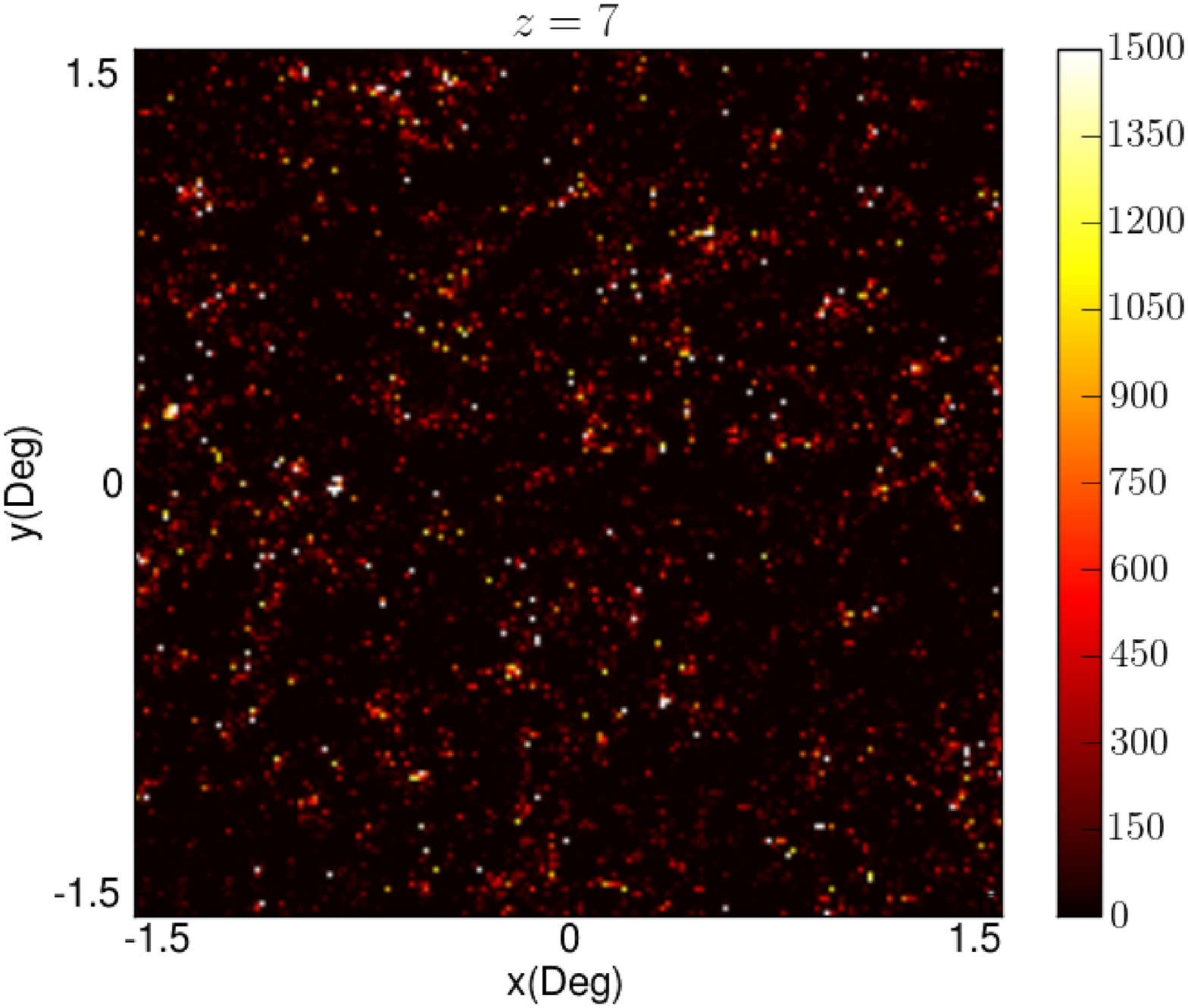}
\includegraphics[scale=0.34]{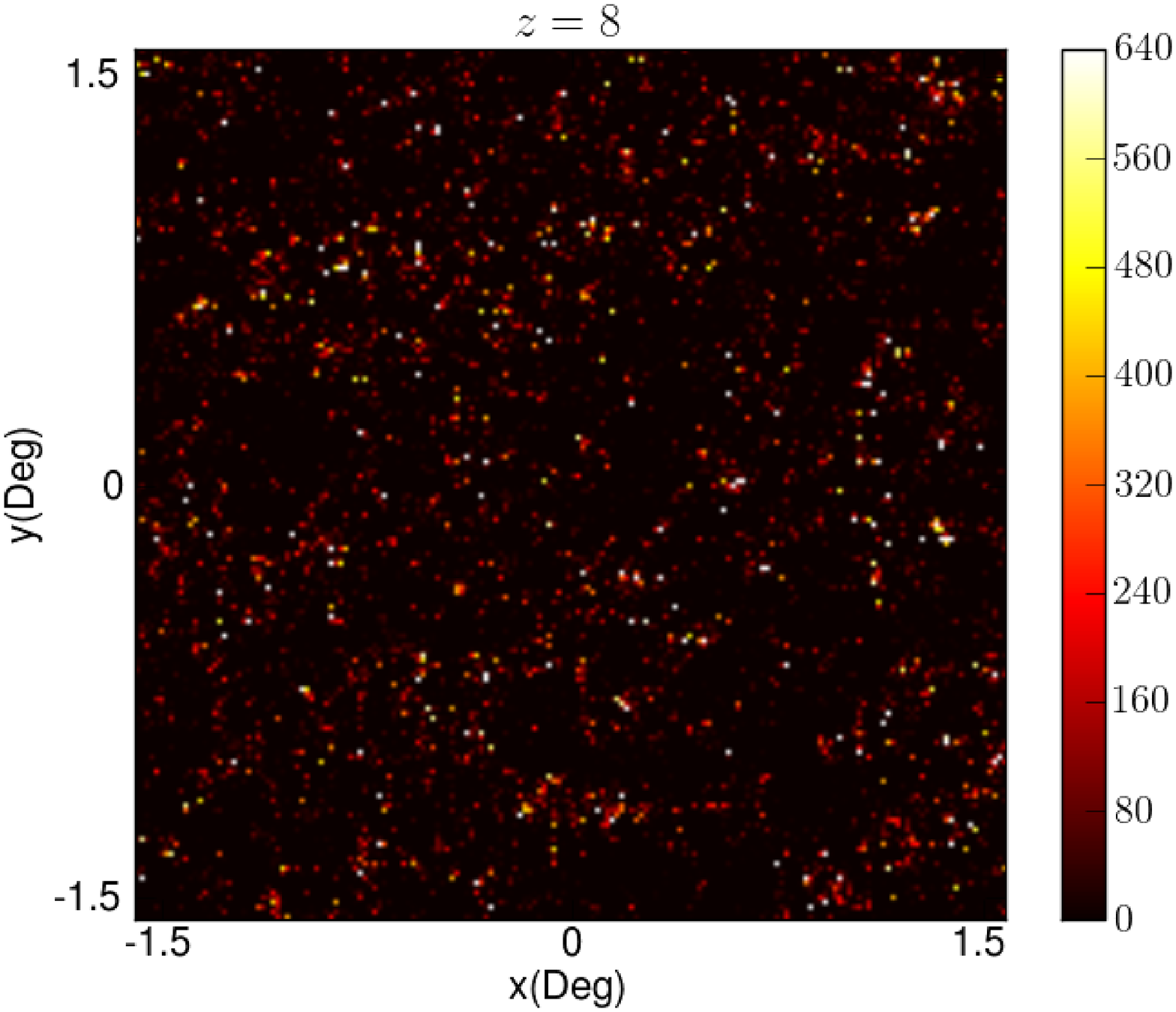}
\caption{The sky maps of the CII emission at $z=6$, 7 and 8 from top to bottom, respectively using \cite{DeLucia07} simulations to calculate the CII line intensities. 
Each of these maps span 3 degrees in each direction and the color bar to the right show the intensity scaling in units of Jy/sr. 
}\label{fig:CII_skymap}
\end{figure}

\section{CO contamination to CII line intensity variations}

When attempting to observe the $z > 6$ CII line intensity variations, the same observations will become sensitive to variations in the intensity of
other emission lines along the line of sight. In particular low-redshift variations will be imprinted by a variety of CO transitions.
This foreground line contamination distorts the CII signal and introduces additional corrections to the measured power spectrum, which will not be due to CII alone.

\begin{figure}[htb]
\includegraphics[scale = 0.43]{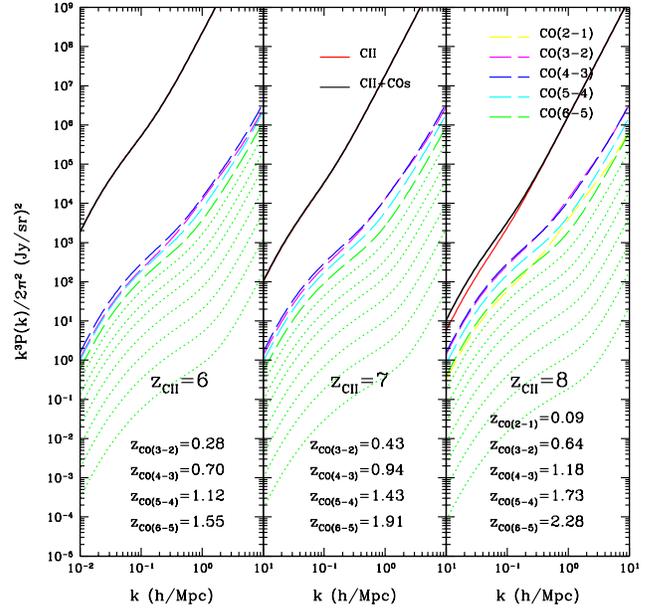}
\caption{\label{fig:P_CII_CO} The contaminated CII  total power spectrum $P^{\rm cont}_{\rm CII}$ 
(black solid lines) at $z=6$, $z=7$ and $z=8$. 
The CII total power spectrum $P_{\rm CII}^{\rm tot}$ is
calculated from the hot gas in the simulation (red solid line) 
(note that the 
$P^{\rm cont}_{\rm CII}$ and $P_{\rm CII}^{\rm tot}$
are almost overlapped at $z=6$ and $z=7$ because of the relatively 
smaller CO line contamination). The other long dashed and green dotted 
lines are calculated with the $L_{\rm CO(m-n)}$ 
given in \cite{Visbal10}. The green dotted lines from upper to 
lower are CO(7-6), CO(8-7) ... CO(13-12) respectively.}
\end{figure}

As an example here we will focus on the contamination from the CO lines as we expect those intensities to be of the same order of magnitude as the CII line intensity.
Given the difference in the CII rest frequency and those of the CO spectrum for each of the transitions, for a given redshift targeting CII line emission,
the frequency of observations and the bandwidth of the observations will correspond to a certain low redshift range in which contaminating CO contribution will be present.
Here we calculate  the CO power spectrum following \cite{Visbal10} with a 
CO luminosity formula as the function of the redshift and 
halo mass \citep{Visbal10}
\ba
L_{\rm CO(m-n)}&=&6.6\times 10^6\bigg( \frac{R_{\rm line}}{3.8\times 10^6}\bigg)\bigg( \frac{M}{10^{10}M_{\odot}} \bigg) \\ \nonumber
&\times& \bigg(\frac{1+z}{7}\bigg)^{3/2}\frac{f_*}{\epsilon_{\rm duty}}L_{\odot},
\ea
where $R_{\rm line}$ is the ratio of star formation rate and the luminosity for a 
emission line, $f_*=0.1$ is the fraction of gas in a halo that can form 
stars, and $\epsilon_{\rm duty}=0.1$ is the duty cycle which is canceled
when computing the CO intensity in \cite{Visbal10}.

\begin{figure*}[htb]
\centerline{
\includegraphics[scale = 0.45]{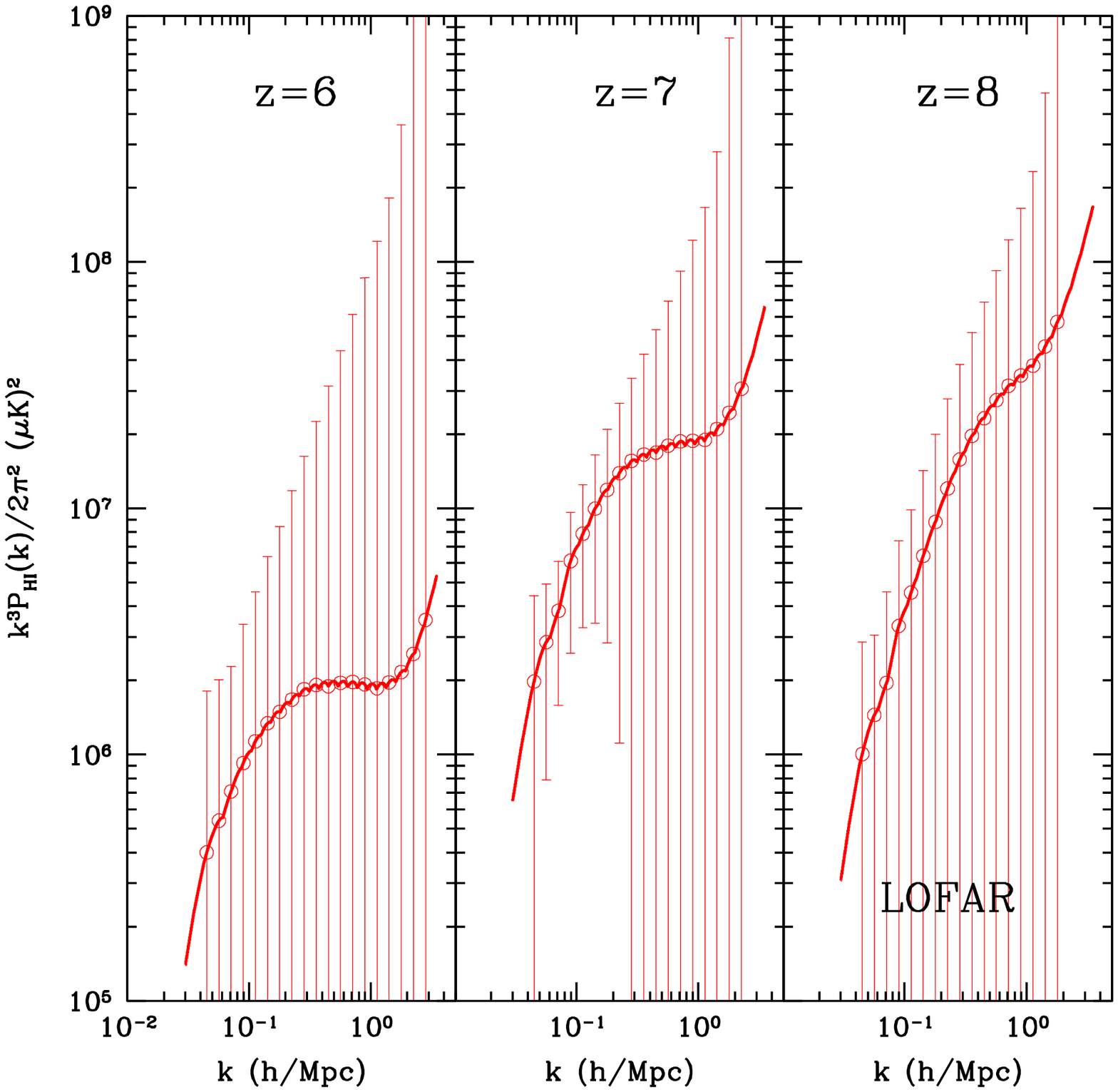}
\includegraphics[scale = 0.45]{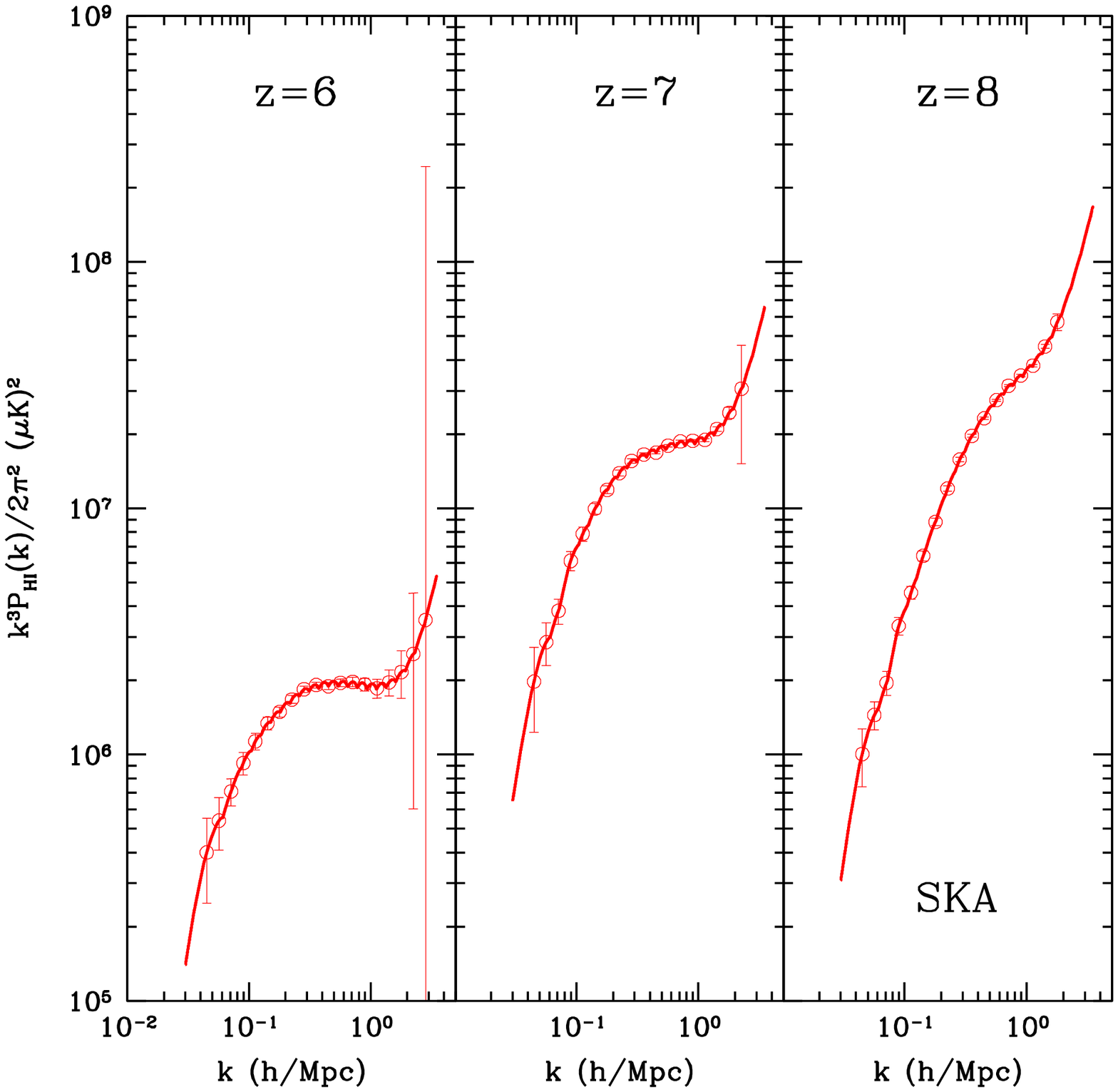}
}
\caption{\label{fig:P_HI} The power spectrum of 21-cm emission
at $z=6$, 7 and 8. The error bars in the left panel are estimated
with a setup similar to LOFAR while the ones in the right
panel are from a setup similar to SKA.}
\end{figure*}

We find this formula has some deviations from the results of our previous simulations 
\citep{Gong11}, but it is still a good approximation when considering the halo mass 
range $10^8 < M < 10^{13}\ \rm M_{\odot}/h$ in which we perform the calculation. 
The advantage of this formula is that we can calculate the luminosity of 
the CO lines at an arbitrary halo mass and redshift. Also, note that take assume $L_{\rm CO}\sim M$ instead of
$L_{\rm CO}\sim M_{\rm CO}$ when calculating the large-scale structure bias factor of CO emitting galaxies.

We show how the CO contamination for CII intensity measurements
from a variety of redshift ranges for $J=2-1$ to $J=13-12$ transitions of CO combine
in Fig.~10. The frequency of the CII line which is emitted at high redshift, e.g. $z=7$, will be
redshifted to $\nu_0=\nu_{\rm CII}/(1+z)\simeq 237\ \rm GHz$ at present. 
As shown in Fig.~\ref{fig:P_CII_CO} 
the main contamination is from the first five CO lines, CO(2-1) to CO(6-5),
and the contamination of the CO lines provide about $2\%$ and
$30\%$ to the total intensity power spectrum at $z=7$ 
and $z=8$, respectively, for large scales ($k<1\ \rm h/Mpc$). 
The black solid lines shown in the plot are the contaminated CII total power 
spectrum for the hot gas case of our simulation, which is the sum 
of the CII total power spectrum $P^{\rm tot}_{\rm CII}(k)$ 
(red solid line) and the CO total power spectrum $P^{\rm tot}_{\rm CO}(k)$
(dashed and dotted lines). 
At small scales ($k>1\ \rm h/Mpc$), the shot noise becomes the dominant
component in the power spectrum, and we find the shot noise of the CII
emission is generally greater than that of the CO emission. We note that
this result is likely subject to assumptions on the CO luminosity
 caused by the CO luminosity we use, since the CO luminosity and especially the CO luminosity-halo mass relation.
We find a weaker dependence on the halo mass than \cite{Visbal10} in our CII emission model.

The other atomic fine-structure emission lines at longer wavelengths than CII,
for example NII at 205 $\mu m$ and CI with 370 and 609 $\mu m$ can also contaminate 
CII emission. The ratios of $L_{\rm CII}/L_{\rm NII}$ and $L_{\rm CII}/L_{\rm CI}$ are about $8\sim10$ according to the observations 
\citep{Malhotra01,Oberst06}, while their luminosities are either comparable or slightly
lower than the CO luminosity. A CII study of reionization will also involve cross-correlation studies at low redshifts to untangle
various line contaminations. In a future paper we will return to such a study using a numerical setup of a CII experiment of the type we propose here.

\begin{figure*}[!t]
\centerline{
\includegraphics[scale = 0.45]{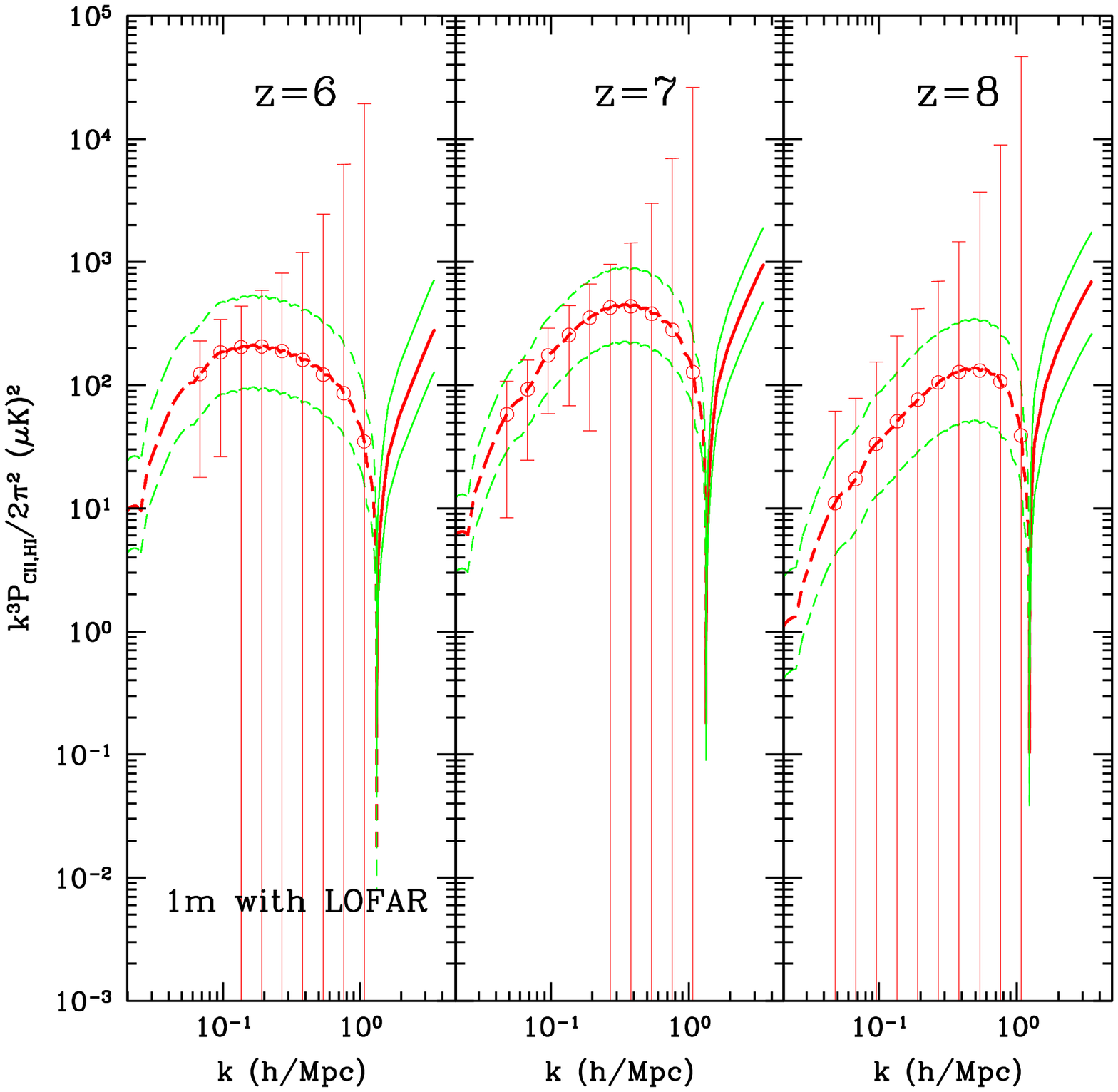}
\includegraphics[scale = 0.45]{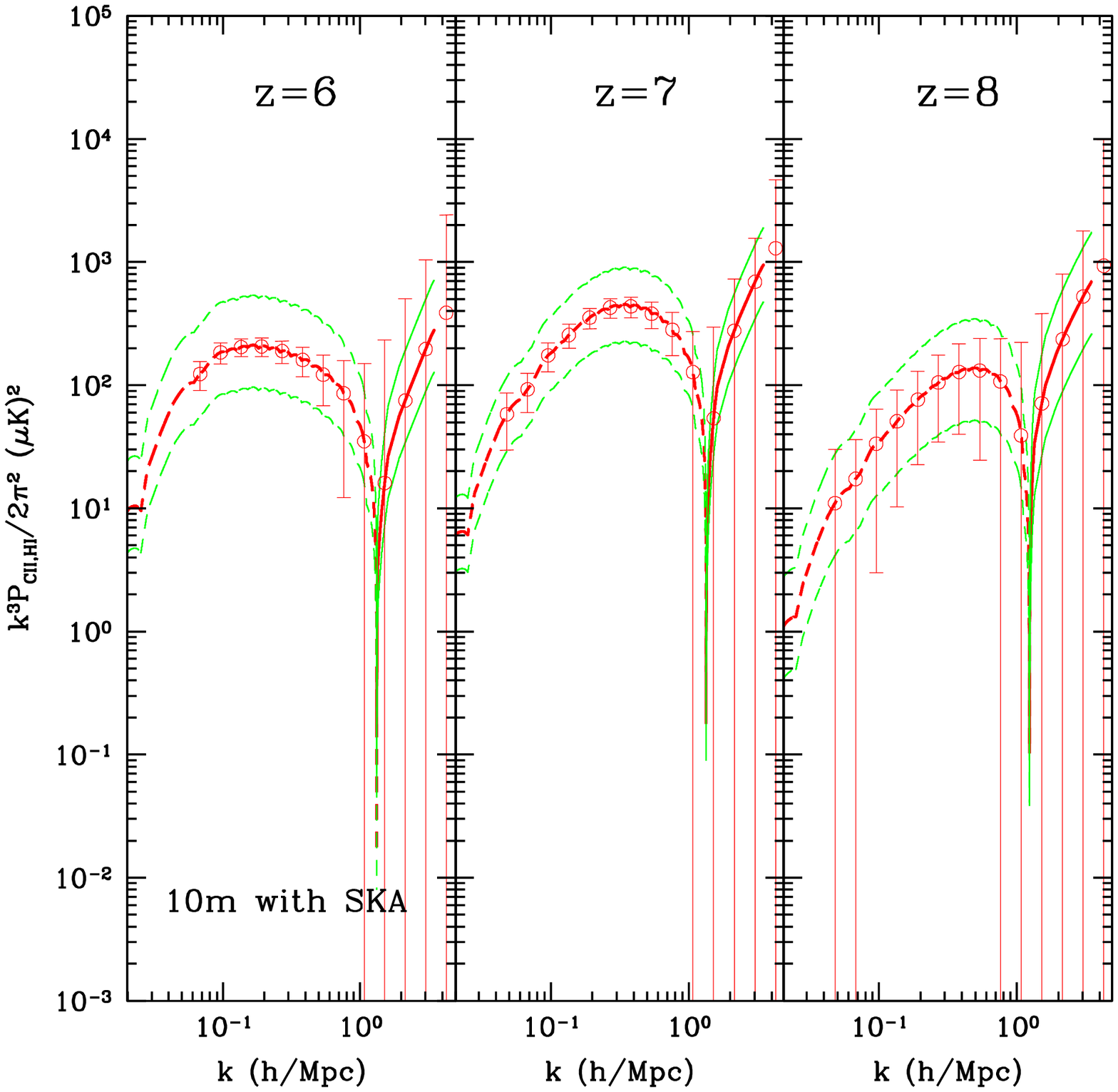}
}
\caption{\label{fig:P_CII_HI} The cross
power spectrum of the CII and the 21-cm emission line at 
$z=6$, $z=7$ and $z=8$. The red dashed lines denote the negative 
correlations while the red solid lines denote positive correlations.
The $1\sigma$ errors are also shown in thin red lines.
Here we just consider the 21-cm signal
from the IGM, since the 21-cm signal from neutral gas in galaxies is about 
$10^{-4}$ of the IGM signal \citep{Gong11}. The error bars in the left panel
are estimated using the (sub-) milimeter survey with a 1 m aperture
for CII line with a setup similar to LOFAR for 21-cm measurements,
while 10 m aperture for CII line with a setup similar to SKA for 
21-cm measurements in the right panel.
}
\end{figure*}

\section{Cross-correlation studies between CII and 21-cm Observations}

Since the above described CO contamination lines come from different redshifts
it is necessary that any CII mapping experiment be considered with another tracer of the same high redshift universe.
In particular, low-frequency radio interferometers now target the $z >6$ universe by probing the neutral hydrogen distribution via the 21-cm spin-flip transition.
Thus, we consider the cross correlation of the CII line and the 21-cm emission
at the same redshift to eliminate the  low-redshift contamination, foregrounds, and to obtain a combined probe of the high redshift universe with
two techniques that complement each other. We expect a strong correlation between the CII and 21-cm lines 
because they both trace the same underlying density field and such a cross-correlation will be insensitive to CO lines
since they are from different redshift which depart far away from each other. There could still be lower order effects, such as due to radio foregrounds.
For example, the same galaxies that are bright in CO could also harbour AGNs that are bright in radio and  be present as residual point sources in the 21 cm data. 
Then the cross-correlation will contain the joint signal related to low redshift CO emission and the residual point source radio flux from 21-cm observations.
The issue of foregrounds and foreground removal for both CII alone and joint CII and 21-cm studies are beyond the scope of this paper. We plan to return to
such a topic in an upcoming paper (Silva et al. in preparation).

Here we calculate the power spectra for the CII-21cm correlation using the correlation between the matter density field and the 21cm brightness temperature obtained from a simulation made using the Simfast21 code \citep{Santos10}, with the further modifications described in \citep{Santos11} to take into account the unresolved halos.
This code uses a semi-numerical scheme in order to simulate the 21cm signal from the Reionization Epoch. The simulation generated has boxes with a resolution of {\bf $1800^3$} cells and a size of L=1Gpc. With an ionizing efficiency of 6.5 we have obtained the mean neutral fraction fraction of 0.05, 0.35, and 0.62 
and an average brightness temperature of 0.63 mK, 6.44 mK, and 14.41 mK 
for $z=6$, $z=7$, and $z=8$ respectively. The power spectrum of 21-cm emission is also shown in Fig.~\ref{fig:P_HI},
the blue dashed error bars are estimated from the LOFAR and the 
red solid ones are from the SKA (see Table~2 for experimental parameters).

\begin{figure*}[!t]
\centerline{
\includegraphics[scale = 0.45]{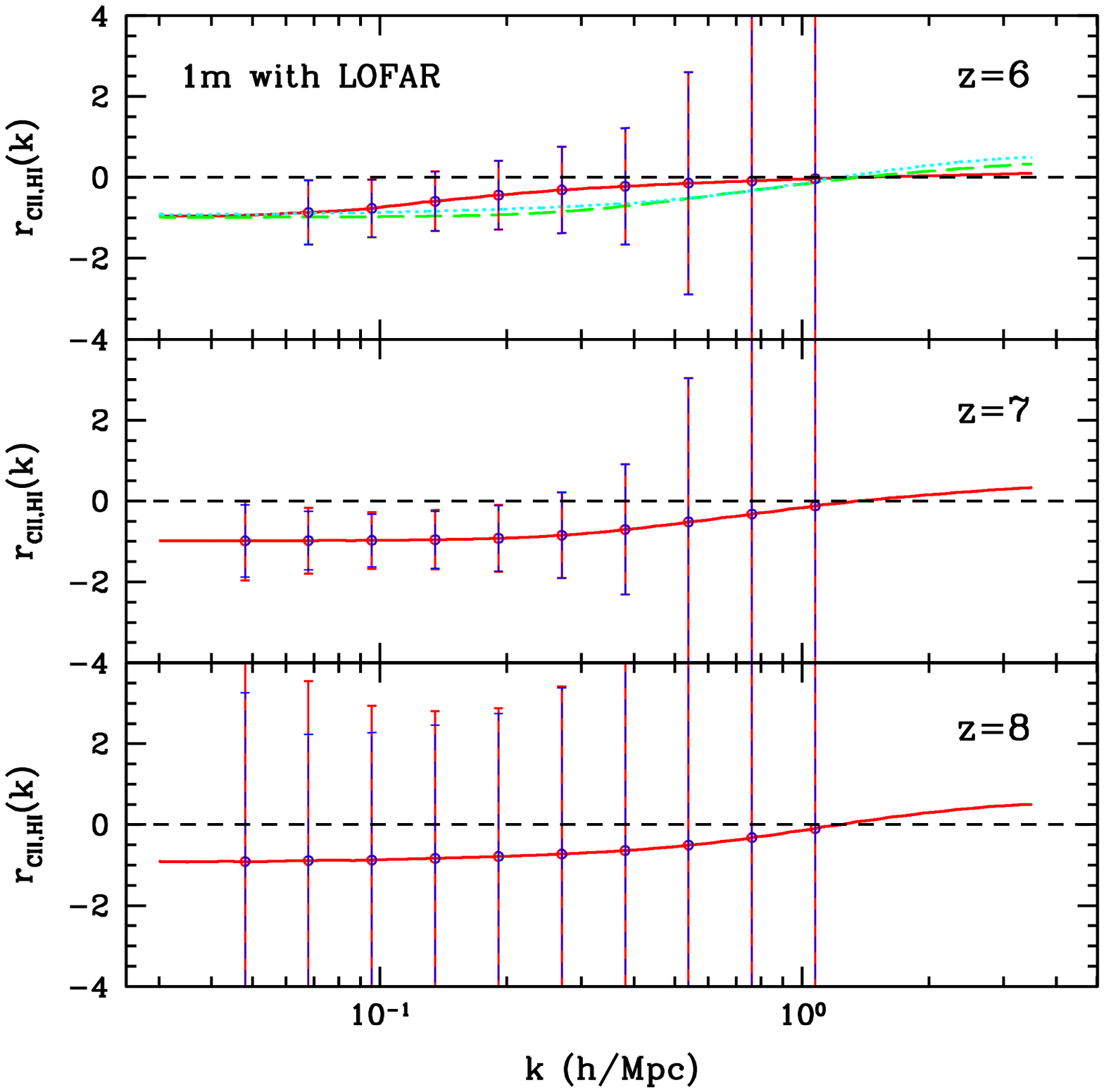}
\includegraphics[scale = 0.45]{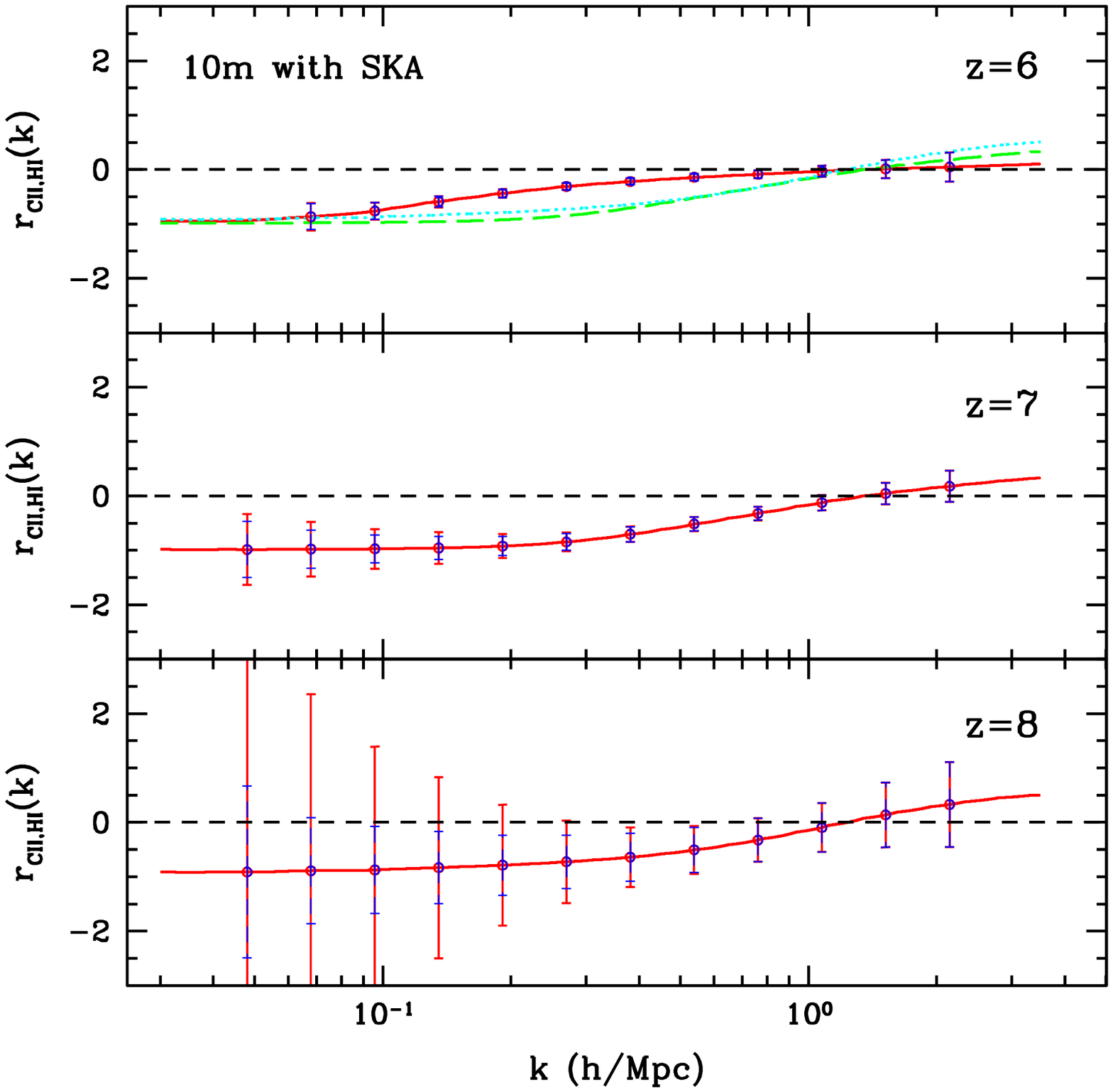}
}
\caption{\label{fig:r} The cross-correlation coefficient of
the CII and 21-cm emission for 1m and 10m aperture at $z=6$, 
$z=7$ and $z=8$. The error bars of r are also shown 
(red solid), and the blue dashed ones are the contribution from
the 21-cm emission with the LOFAR (left panel), and with
the SKA (right panel). We find the 21-cm noise 
dominates the errors at $z=6$ and 7. In the two top panels, the 
$r_{\rm CII,HI}(k)$ at $z=7$ (green dashed) and $z=8$ (cyan dotted) are
also shown to denote to show the evolution of the ionized bubble size at these redshifts relative to $z=6$.
As expected, the bubble size is greater at $z=6$ than that at $z=7$ and 8.}
\end{figure*}

In Fig.~\ref{fig:P_CII_HI}, we show the cross power spectrum of the CII
and 21-cm emission line (red thick lines) and $1\sigma$ uncertainty 
(red thin lines) at $z=6$, $z=7$ and $z=8$. The error bars of the cross power
spectrum are obtained by the assumed milimeter spectrometeric survey with 
1 m and 10m aperture for CII line 
and LOFAR (left panel) and SKA (right panel) for 21-cm emission.
Note that the correlation is negative on large scales (small k) when the 
correlation between the ionization fraction and the matter density dominates 
and positive on small scales when the matter 
density auto-correlation is dominating.
This can be seen by looking at the expression for the cross-power spectrum, which to linear order is given by
\be \label{eq:P_CII_HI}
P_{\rm CII,HI}(z,k) \propto 4/3(1-\bar{x}_{\rm i})P_{\delta \delta}-\bar{x}_{\rm i}P_{\delta_{x}\delta},
\ee
where $\bar{x}_{\rm i}$ is the average ionization fraction and 
$P_{\delta_x \delta}$ is the cross
power spectrum of the ionized fraction and the dark matter.
To show the cross-correlation illustratively, we also plot the 
cross-correlation coefficient at $z=6$, $z=7$ and $z=8$ 
in Fig.~\ref{fig:r}, which is estimated by 
$r_{\rm CII,HI}(k)=P_{\rm CII,HI}(k)/\sqrt{P_{\rm CII}(k)P_{\rm HI}(k)}$.
The typical ionized bubble size can be reflected by the scale that
$r_{\rm CII,HI}(k)$ rises up from $-1$ towards zero. As indicated by
the $r_{\rm CII,HI}(k)$ at $z=6$, 7 and 8 in the top panels of Fig.~\ref{fig:r},
the ionized bubble size at $z=6$ is greater than that at $z=7$ and 8
which is consistent with the current reionization scenario.

\section{Outline of a CII Intensity Mapping Experiment}

We now discuss the requirements on an instrument designed to measure
the CII line intensity variations and the power spectrum at $z >
6$. Here we provide a sketch of a possible instrument design including
calculated noise requirements based on current technology; a detailed
design study will be left to future work as our concept is further
developed. 

An experiment designed to statistically measure the CII transition at
high redshift requires a combination of a large survey area and low
spectral resolution; because of this, the use of an interferometer
array such as the one suggested in \cite{Gong11} for CO seems
impractical. Instead we opt for a single aperture scanning experiment
employing a diffraction grating and bolometer array in order to
provide large throughput.  A set of useful equations is outlined in the Appendix
to convert from bolometer noise to the noise of the actual intensity power spectrum measurement.

\begin{table*}
\centering                         
\caption{Experimental Parameters for a Possible CII Mapping Instrument.}     
\begin{tabular}{l | c c c}   
\hline\hline                 
   Aperture diameter (m) & 1 & 3 & 10\\
\hline 
   Survey Area ($A_{\rm S}$; deg$^2$) & 16 & 16 & 16\\
   Total integration time (hours) & 4000 & 4000 & 4000\\
   Free spectral range ($B_\nu$; GHz) & $185{-}310$ & $185{-}310$ & $185{-}310$ \\
   Freq. resolution ($\delta_\nu$; GHz) & 0.4 & 0.4 & 0.4 \\
   Number of bolometers & 20,000 & 20,000 & 20,000 \\
   Number of spectral channels & 312 & 312 & 312 \\
   Number of spatial pixels & 64 & 64 & 64 \\
   Beam size$^{\mathrm{a}}$ ($\theta_{\rm beam}$; FWHM, arcmin) & $4.4$ & $1.5$ & $0.4$\\
   Beams per survey area$^{\mathrm{a}}$ & $2.6 \times 10^{3}$ & $2.3 \times 10^{4}$ & $2.6 \times 10^{5}$ \\ 
   $\sigma_{\rm pix}$: Noise per detector sensitivity$^{\mathrm{a}}$ (Jy$\sqrt{\mathrm{s}}$/sr) & $2.5 \times 10^{6}$ & $2.5 \times 10^{6}$ & $2.5 \times 10^{6}$ \\
   $t^{\rm obs}_{\rm pix}$: Integration time per beam$^{\mathrm{a}}$ (hours) & 100 & 11 & 1.0 \\
\hline
$z=6$ $V_{\rm pix}$ (Mpc/h)$^3$ & 217.1 & 24.1 & 2.2 \\
$z=7$ $V_{\rm pix}$ (Mpc/h)$^3$ & 332.9 & 37.0 & 3.3 \\
$z=8$ $V_{\rm pix}$ (Mpc/h)$^3$ & 481.3 & 53.5 & 4.8 \\
   \hline
   $z=6$  $P^{\rm CII}_N$ (Jy/sr)$^2$ (Mpc/h)$^3$ & 5.4$\times10^9$ & 5.4$\times10^9$ & 5.3$\times10^9$\\
   $z=7$  $P^{\rm CII}_N$ (Jy/sr)$^2$ (Mpc/h)$^3$ & 4.8$\times10^9$ & 4.9$\times10^9$ & 4.8$\times10^9$\\
   $z=8$  $P^{\rm CII}_N$ (Jy/sr)$^2$ (Mpc/h)$^3$ & 4.4$\times10^9$ & 4.4$\times10^9$ & 4.3$\times10^9$\\
\hline \hline
\multicolumn{4}{c}{$^{\mathrm{a}}$ values computed at $238 \,$GHz, corresponding to CII at $z=7$.}                                
\end{tabular}
\label{table_parms}     
\end{table*}

For proper sampling of the CII signal on cosmological scales and cross-correlation with 21-cm experiments, we
would require at minimum a survey area of 16 deg$^2$ and a free
spectral range (FSR) of 20 GHz.  At a redshift of $z=7$, a box 4
degrees on a side and 20 GHz deep corresponds to a rectangular box
with a comoving angular width of 443 Mpc/h and a depth along the line
of sight of 175 Mpc/h.  However, since larger FSRs are easily achieved
with diffraction grating architectures and would allow for better
measurement of the reionization signal and separation of its
foregrounds, the instrumental concept presented here covers the $220
\,$GHz atmospheric window with an FSR of $125 \,$GHz.  Concretely,
covering from 185 to 310 GHz with a spectral resolution of $0.4 \,$GHz
allows measurement of CII in the range $5.1 \leq z \leq 9.3$ with a
smallest redshift interval $\Delta z$ of $0.01$.  

The integration time
per beam on the sky required to survey a fixed area depends on a
number of parameters including the size of the aperture, the number of
independent spatial pixels in the spectrometer, and the bandwidth of
each spectral element.  Changing the survey area or bandwidth will
affect the minimum $k$ values probed in the 3-d power spectrum as well
as the number of modes used in the error calculation.  To generate
concrete examples in this discussion, we concentrate on calculating
the CII power spectrum at $z=6$, 7 and 8 and assume that we will make
use of 20 GHz effective bandwidths at frequencies centered at these
redshfits; such effective bands are readily achieved by summing
neighbouring high resolution spectral bins.  At $z=7$, for example, a
20 GHz bandwidth corresponds to a $\Delta z=0.62$.  Averaging over a
larger spectral window to reduce noise will make the cosmological
evolution along the observational window non-negligible.

To understand the effect of spatial resolution on this measurement we
consider three possible primary aperture sizes of 1, 3 and 10 m; the
resulting instrumental parameters are listed in Table 1. These
apertures correspond yield beams of 4.4, 1.5, and 0.4 arcmin and
comoving spatial resolutions of 8.3, 2.8, 0.8 Mpc/h at an observing
frequency of 238 GHz ($z=7$), respectively. These apertures probe the
linear scales ($k<1.0$ h/Mpc) and are well matched to 21-cm
experiments.  In this experimental setup light from the aperture is
coupled to a imaging grating dispersive element illuminating a focal
plane of bolometers, similar to the Z-Spec spectrometer but with an
imaging axis (Naylor et al. 2003; Bradford et al. 2009).  

We assume a fixed spectral resolution of $\sim 500 \,$km/s ($=0.4 \,$GHz as
discussed above) in each of the three cases giving a spatial
resolution of 3.5 Mpc/h and a maximum $k$ mode of $k\approx 0.91$
h/Mpc.  Current technology can allow fabrication of $\sim 20,000$
bolometers in a single focal plane; for the instrument concept we
freeze the number of detectors at this value.  The spectrometric
sensitivity of the grating spectrometer design is estimated using a
model of the instrument which includes loading from the CMB,
atmosphere and instrument contributions assuming realistic
temperatures and emissivities for each.  
 
As an example, using equations in Appendix A, for  a 3 m aperture at $238 \,$GHz and a 
$NEP_{\mathrm{Det}} = 1 \times 10^{-18} \,$W HZ$^{-1/2}$, a value
readily achieved using current detector technology, we find a noise equivalent flux density
 (NEFD) of $0.42 \,$Jy s$^{1/2}$ per pixel per spectral resolution
 element.  This NEFD can then be converted to the units of Jy
 s$^{1/2}$/sr used in Table 1 by multiplication by the solid angle of
 the telescope response.  For observations corresponding to $z =7$,
 this results in a noise per pixel of $2.5 \times 10^{6}$ Jy
 s$^{1/2}$/sr.

These noise power spectra based on Table~1 are  shown for the 1 and 10 meter aperture in Fig.~8.
The proposed CII experiments involve three different apertures at 1, 3, and 10 m, 
but all three options make use of the same spectrometer with a total 
$\sim$ 20,000 bolometers that make up 64 individual pixels on the sky.
The statistical detection shown in figure \ref{fig:P_CII} using the 
instrumental parameters listed in Table~1 can be obtained by noting 
that the error on each of the binned power spectrum measurements
is $\Delta P = (P^{\rm CII}_N+P_{\rm CII})/\sqrt{N_m}$, where 
$P_{\rm CII}$ is the CII power spectrum, including shot-noise and $N_m$ is the number of $k$-modes
available for each power spectrum measurement.

The noise parameters are such that with an integration time of 4000
hours the CII power spectrum at $z =6$ is detected with a
signal-to-noise ratio, ${\rm S/N}=\sqrt{\sum_{\rm bins}
  {\left(P(k)\over \Delta P(k)\right)^2}}$, of 12.6, 13.2, and 13.5
for 1, 3, and 10 m aperture cases, respectively.  The corresponding
values for $z=7$ are 2.2, 2.2, and 2.3 respectively, while at $z=8$ the
signal-to-noise ratio values are less than 1 for all three options.

For the cross-correlation with 21-cm data, we assume that observations will be
conducted in overlapping areas on the sky by both the CII and the
21-cm experiment.  This is a necessary feature of cross-correlation
study leading to some coordination between two experiments.  We reduce
both data sets to a common, and a lower resolution cube in frequency
and count the total number of common modes contributing to a given bin
in $k$ space using the same method for noise calculation as in
\cite{Gong11}. The error in a given $k$ bin can be written as
$\sqrt{(P_{\rm CII,H}^2+P_{\rm CII}P_{\rm H})/(2 N_m(k))}$, where $N_m(k)$ is
the number of modes falling in that bin.

When describing the 21-cm observations we assume sensitivities similar to
LOFAR with parameters $A_{\rm tot}=2.46\times10^4$ m$^2$ and $T_{\rm sys}=449$K at 150 MHz, bandwidth
of 12 MHz, and resolution of 0.25 MHz (see Table~2 for SKA parameters).  
Here, we further assume that the baseline density distribution for LOFAR and SKA is constant on the 
u-v plane up to a maximum baseline $\rm D_{max}$. This is a reasonable 
approximation for the dense central core of these experiments which is 
the only configuration we're using for the measurements.
The cross-correlation is detected with a signal-to-noise ratio of 3 to 4 at $z=6$ and 4 to 5 at $z=7$ when any of the three CII options is combined with LOFAR.
There is a factor of a 3 to 4 improvement when combined with SKA leading to signal-to-noise ratios of 10, 11 and 12 for the $z=6$ cross-correlation
with 1, 3, and 10 m options, respectively. At $z=7$, the signal-to-noise ratios are 10, 12, and 13, while at $z=8$ they are 3, 4 and 5, respectively.
While the CII power spectrum cannot be detected at $z=8$ using any of the three options in Table~1, 
the cross-correlations can be detected when one of the proposed CII experiments is combined with SKA.

The cross-correlation coefficient between CII and 21-cm (Fig.~\ref{fig:r})
shows a scale that $r_{\rm CII, HI}(k)$ begins to rise from -1 to 0 due to the size of
the ionized bubbles surrounding galaxies during reionization
(Eq.~\ref{eq:P_CII_HI}). This scale captures both
the size of the typical ionized bubbles and the mean value of the
ionization fraction $\bar{x}_i$. The observational measurement of this
rise-up scale is challenging for a first-generation
CII experiment involving 1 or 3 m aperture combined with a first-generation 21-cm experiment like
LOFAR, but is likely to be feasible with an improved CII experiment combined with SKA.

\begin{table}[!t]
\centering
\caption{Sensitivities of LOFAR and SKA 21-cm experiments at 150MHz.}
\begin{tabular}{l | c | c}
\hline\hline
   Instrument & LOFAR & SKA \\
\hline
   FoV (deg$^2$) & 25 & 25\\
   Bandwidth (MHz) & 12 & 12\\
   Freq. resolution (MHz) & 0.25 & 0.25 \\
   System temperature (K) & 449 & 369 \\
   maximum baseline (m)& 2000 & 5000\\
   Total integration time (hours) & 1000 & 4000 \\
   total collecting area at 150 MHz (m$^2$) & 2.46$\times10^{4}$ & 1.4$\times10^{6}$\\
\hline
   $z=6$  $P^{\rm 21cm}_N$ [K$^2$ (Mpc/h)$^3$] & 7.8$\times10^{-1}$ & 1.0$\times10^{-3}$ \\
   $z=7$  $P^{\rm 21cm}_N$ [K$^2$ (Mpc/h)$^3$] & 1.0 & 1.5$\times10^{-3}$ \\
   $z=8$  $P^{\rm 21cm}_N$ [K$^2$ (Mpc/h)$^3$] & 1.4 & 2.3$\times10^{-3}$ \\
\hline \hline
\end{tabular}
\label{tab:21setup}
\end{table}

\section{Conclusion and Discussion}

In this paper, we have estimated the mean intensity and the spatial intensity 
power spectrum of the  CII emission from galaxies at $z > 6$. We first calculated the CII intensity analytically for
both the ISM of galaxies and the diffuse IGM  and showed that the CII emission from dense gas in the ISM of galaxies
is much stronger than that from the IGM. Then, to check the analytical calculation,
 we used the hot gas or the hot+warm gas from a simulation to find the CII mass in a halo,
and further calculated the CII number density and intensity. We found
that the two methods are in good agreement especially
at high redshifts we are interested in. Next, we computed the CII clustering power spectrum
assuming the CII luminosity is proportional to the CII mass, 
$L_{\rm CII}\sim M_{\rm CII}$. We compared our CII power spectrum with
that derived from the $L_{\rm CII}$-$L_{\rm CO(1-0)}$ relation, and
found that they are consistent in the $1\sigma$ level. We also explored the
contamination of the CII emission by the CO lines at lower redshift,
and found the contamination can lead to $2\%$ and $30\%$
enhancement of the CII power spectrum at $z=7$ and $z=8$. 

To reduce the foreground contamination and to
improve our scientific understanding of reionization we propose here a cross-correlation study between
the CII and 21-cm emission in the overlapping redshift ranges and the same part of the sky.
The cross-correlation exists since they both trace the same 
matter distribution. At large scales the correlation is due to ionized bubbles surrounding CII bright galaxies
while at smaller scales both CII galaxies and ionized bubbles trace the underlying density field.
We have outlined 3 potential CII experiments using 1, 3 and 10 m aperture telescope outfitted with a bolometer array spectrometer
with 64 independent spectral pixels. A 1 or 3 meter aperture CII experiment is matched to
a first generation 21-cm experiment such as LOFAR while an improved CII experiment can be
optimized to match a second generation 21-cm experiment like SKA. 
We find that the overall ability to extract details on reionization requires a careful coordination
and an optimization of both CII and 21-cm experiments. We have not discussed the issues related to foregrounds, including
AGN-dominated radio point sources that dominate the low-frequency observations and dusty galaxies that dominate the high-frequency CII
observations and Galactic foregrounds such as dust and synchrotron. In future papers we will return to these topics and
also expand our discussion related to the instrumental concept as we improve the existing outline to an actual design.

\begin{acknowledgments}
We thank participants of the Keck Institute for Space Studies' (KISS) Billion Years 
workshop for helpful discussions. This work was supported by NSF CAREER AST-0645427 
and NASA NNX10AD42G at UCI. 
MGS and MBS acknowledges support from FCT-Portugal under grant PTDC/FIS/100170/2008.
\end{acknowledgments}



\appendix

In this Appendix we summarize a set of key equations necessary to obtain the final noise power spectrum of the intensity fluctuation measurements.
We focus here on a single aperture scanning experiment with a bolometer array, but intensity fluctuation measurements can also be pursued with
interferometeric measurements (especially 21-cm and CO). In the context of 21-cm cross-correlation,
we also provide the interferometer noise formula below without a detailed derivation.

Unlike the case of CMB, where intensity fluctuation measurements are two-dimensional on the sky with the noise given in \cite{Knox95}, the
CII line intensity fluctuation measurements we pursue here are  three-dimensional with information related to the spatial inhomogeneities 
both in angular and redshift space. For simplicity, we ignore redshift space distortions here and only consider a case where spatial variations
are isotropic, thus the three-dimensional power spectrum is directional independent. The line of sight mode $k_\parallel$ and angular/transverse mode $k_\perp$
are related to each of the $k$-mode via, $k_\parallel=\mu k$, $k_\perp=\sqrt{k^2-k_\parallel^2}$, as a function of the angle in Fourier space $\mu$.

The noise power spectrum of intensity fluctuations take the form
\begin{equation}
P^{\rm CII}_N(k,\mu) = V_{\rm pix} \frac{\sigma_{\rm pix}^2}{t^{\rm obs}_{\rm pix}} {\rm e}^{(k_\parallel/k_1)^2+(k_\perp/k_2)^2}
\end{equation}
where $\sigma_{\rm pix}$ is the noise per detector sensitivity,
$t^{\rm obs}_{\rm pix}$ is the integration time per beam/pixel, and $V_{\rm pix}$
is the pixel volume in real space. The exponential factor captures both the spatial ($k_2$) and radial ($k_1$) resolution. The spatial resolution is
set by the instrumental beam while the resolution along the radial direction are set by the spectral bin $\delta_\nu$. These are respectively,
$k_2(z)= 2\pi/(r(z)\theta_{\rm beam})$ and $k_1(z)=H(z)\nu/[c(1+z)\, \delta_\nu]$, where $r(z)$ is the comoving radial distance and $H(z)$ is the Hubble parameter.
The former can be obtained from the integral $r(z)=\int_0^z c/H(z)\,dz$.
For reference, $k_1\simeq0.29 (5\, {\rm arcmin}/\theta_{\rm beam})\, \rm (h/Mpc)$ and $k_2\simeq0.69(400\, {\rm MHz}/\delta_\nu)\, \rm (h/Mpc)$  at $z=7$.

The total variance of the power spectrum  is then
\begin{equation}
{\rm var}[P_{\rm CII}(k,\mu)] = \left[P_{\rm CII}(k) + P^{\rm CII}_N(k,\mu) \right]^2 \, ,
\end{equation}
where the first team denotes the usual cosmic variance.
In the case where a spherically averaged  power spectrum measurement is pursued, one can simply take the minimum variance estimate given the above $\mu$-dependent
noise associated with the exponential  factor. Assuming $\sigma_{\rm pix}$ is $\mu$-independent, the $\mu$-averaged variance of the power spectrum meausrement is then
simply (Lidz et al. 2011)
\begin{equation}
{\rm var}[\bar{P}_{\rm CII}(k)] = \frac{\left[P_{\rm CII}(k) + \bar{P}^{\rm CII}_N(k) \right]^2}{N_m(k)} \, ,
\end{equation}
where $N_m(k)$ is the number of total modes that leads to the power spectrum measurement at each $k$
and 
\begin{equation}
\bar{P}^{\rm CII}_N(k) = V_{\rm pix} \frac{\sigma_{\rm pix}^2}{t^{\rm obs}_{\rm pix}}  \, .
\end{equation}

The number of modes at each $k$ for the $P(k)$ measurement is
\be
N_m(k) = 2\pi k^2\Delta k \frac{V_{\rm S}}{(2\pi)^3} \, ,
\ee
where $\Delta k$ is the Fourier bin size and $(2\pi)^3/{V_{\rm S}}$ 
is the resolution in Fourier space. Note that the result is obtained
by integrated the positive angular parameter only (i.e. $0<\mu={\rm cos}(\theta)<1$). 

The total survey volume is ${V_{\rm S}}=r(z)^2yA_{\rm S}B_\nu$, where $A_{\rm S}$ is the survey area (in radians), 
$B_\nu$ is the total bandwidth of the measurement,  $y=\lambda (1+z)^2/H(z)$ is the factor to convert the frequency 
intervals to the comoving distance at the wavelength $\lambda$.
A useful scaling relation for $V_{\rm S}$ for CII measurements is
\be
V_{\rm S}(z) = 3.3\times 10^7 {\rm (Mpc/h)^3} \left(\frac{\lambda}{158\ \mu m}\right) \sqrt{\frac{1+z}{8}}\left(\frac{A_{\rm S}}{16\ \rm deg^2}\right)\left(\frac{B_\nu}{20\ \rm GHz}\right),
\ee
where $\lambda$ is the emission line wavelength in the rest frame.

In $\bar{P}^{\rm CII}_N(k)$, the volume surveyed by each pixel is $V_{\rm pix}=r^2yA_{\rm pix}\delta_\nu$. Here $A_{\rm pix}$ is the spatial
area provided by the beam, or an individual pixel depending on the instrument design (in radians). 
A simple scaling relation for $V_{\rm pix}$ as a function of the redshift $z$ is
\be
V_{\rm pix}(z) = 1.1\times 10^3 {\rm (Mpc/h)^3} \left(\frac{\lambda}{158\ \mu m}\right)  \sqrt{\frac{1+z}{8}} \left(\frac{\theta_{\rm beam}}{10\ \rm arcmin}\right)^2\left(\frac{\delta_\nu}{400\ \rm MHz}\right).
\ee

{\rm Bolometer noise formula:} To obtain $\sigma_{\rm pix}$, for the CII case with a bolometer array, we
make use of a calculation that estimates
the noise equivalent power of the background ($NEP_{\mathrm{BG}}$) via the quadrature sum
of contributions from Phonon, Johnson, Shot and Bose noise terms using
these loadings and estimates of realistic instrument parameters from
existing experiments (e.g.~Brevik et al.~2010).  Similarly, the total
NEP$_{\mathrm{tot}}$ of the instrument is the square root of the
quadrature sum of $NEP_{\mathrm{BG}}$ and a detector with
$NEP_{\mathrm{Det}} = 1 \times 10^{-18} \,$W HZ$^{-1/2}$, a value
readily achieved using current detector technology.
NEP$_{\mathrm{tot}}$ does not depend on the size of the telescope
aperture; to convert this to the corresponding noise equivalent flux
density (NEFD) in Jy s$^{1/2}$ we use
\begin{equation}
\mathrm{NEFD} = \frac{\mathrm{NEP}_{\mathrm{tot}}}{\eta_{\mathrm{sys}} A \sqrt{2
    \Delta \nu} \exp(-\eta_{\mathrm{sky}})}
\end{equation}
 where $A$ is the area of the telescope aperture.  The numerical values obtained are in Table~1.

{\it 21-cm interferometer noise formula:}  When estimating the cross-correlation, for simplicity, we assume that 
the noise power spectrum amplitude $P^{\rm 21cm}_N$ is a constant and it takes the form (e.g., Santos et al. 2010)
\be
P^{\rm 21cm}_N(k,\theta)=D_A^2 y(z) \frac{\lambda^4 T_{sys}^2}{A_{\rm tot}^2 t_0  n\left(D_A k \sin(\theta)/2\pi\right)},
\ee
where $D_A$ is the comoving angular diameter distance, $A_{\rm tot}$ is the collecting area for one element of the interferometer, $t_0$ is the total integration time 
and the function $n()$ captures the
baseline density distribution on the plane perpendicular to the line of sight, assuming it is already
rotationally invariant ($k$ is the moduli of the wave mode $\mathbf{k}$ and $\theta$ is the angle between $\mathbf{k}$ and the line of sight) \citep{Santos11,Gong11}.
The values are LOFAR and SKA are tabulated in Table~2.


\begin{thebibliography}{61}

\bibitem[Aguirre \& Schaye(2005)]{Aguirre05}
Aguirre, A. \& Schaye, J. 2005, pgqa. conf. 289

\bibitem[Arnett(1996)]{Arnett96}
Arnett, D. 1996, Supernovae and Nucleosynthesis, Princeton, New Jersey: Princeton University Press

\bibitem[Basu et al.(2004)]{Basu04}
Basu, K., Hernandez-Monteagudo, C. \& Sunyaev, R., A. 2004, A\&A., 416, 447


\bibitem[Bock et al.(1993)]{Bock93}
Bock, J. J., Hristov, V. V., Kawada, M. et al. 1993, ApJ, 410, L115

\bibitem[Boselli et al.(2002)]{Boselli02}
Boselli, A., Gavazzi, G., Lequeux, J. \& Pierini, D. 2002, A\&A, 358, 454

\bibitem[Bouwens et al.(2008)]{Bouwens08}
  Bouwens, R. J. et al. 2008, ApJ, 686, 230

\bibitem[Bradford et al.(2009)]{Bradford09}
Bradford, C. M., et al. 2009, ApJ, 705, 112

\bibitem[Breuck et al.(2011)]{Breuck11}
Breuck, C. D., et al. 2011, arXiv:1104.5250

\bibitem[Brevik et al.(2010)]{Brevik10} Brevik, J.~A., et al.\ 
2010, \procspie, 7741, 41  

\bibitem[Carilli(2011)]{Carilli11}
Carilli, C. L. 2011, ApJ, 730, L30

\bibitem[Crawford et al.(1985)]{Crawford85}
Crawford, M. K., Genzel, R., Townes, C. H. \& Watson, D. M. 1985, \apj, 291, 755

\bibitem[Dalgarno \& McCray(1972)]{Dalgarno72}
Dalgarno, A. \& McCray, R. 1972, ARA\&A, 10, 375

\bibitem[Datta et al.(2010)]{Datta10}
Datta, A, Bowman, J. D. \& Carilli, C. L. 2010, \apj, 724, 526

\bibitem[De Looze et al.(2011)]{DeLooze11}
  De Looze, I., Baes, M., Bendo, G. J., Cortese, L., Fritz, J. 2011, MNRAS in press, arXiv.org:1106.1643

\bibitem[De Lucia \& Blaizot(2007)]{DeLucia07}
De Lucia, G. \& Blaizot, J. 2007, \mnras, 375, 2


\bibitem[Field(1958)]{Field58}
Field, G. B. 1958, Proc. I. R. E., 46, 240

\bibitem[Fukugita \& Kawasaki(1994)]{Fukugita94}
Fukugita, M. \& Kawasaki, M. 1994, \mnras, 269, 563


\bibitem[Giallongo et al.(1997)]{Giallongo97}
Giallongo, E., Cristiani, S., D'Odorico, S., Fontana, A. \& Savaglio, S. 1997, seim. proc. 127

\bibitem[Gong et al.(2011)]{Gong11}
Gong, Y., Cooray, A., Silva, M. B., Santos, M. G. \& Lubin, P. 2011, \apj, 728, L46

\bibitem[Haiman et al.(2000)]{Haiman00}
Haiman, Z., Abel, T. \& Rees, M. 2000, \apj, 534, 11

\bibitem[Hernandez-Monteagudo et al.(2006)]{Hernandez06}
Hernandez-Monteagudo, C., Haiman, Z., Jimenez, R. \& Verde, L. 2006, arXiv:0612363

\bibitem[Hernandez-Monteagudo et al.(2008)]{Hernandez07}
Hernandez-Monteagudo, C., Haiman, Z., Verde, L. \& Jimenez, R. 2008, \apj, 672, 33

\bibitem[Keenan et al.(1986)]{Keenan86}
Keenan, F. P., Lennon, D. J., Johnson, C. T. \& Kingston, A. E. 1986, \mnras, 220, 571

\bibitem[Kistler et al.(2009)]{Kistler09}
Kistler, M. D., et al. 2009, ApJ, 705, L104

\bibitem[Knox(1995)]{Knox95}
Knox, L. 1995, \prd, 52, 4307

\bibitem[Komatsu et al.(2011)]{WMAP7}
Komatsu, E., et al. 2011, \apjs, 192, 18

\bibitem[Kramer et al.(2010)]{Kramer10}
Kramer, R. H., Haiman, Z. \& Madau, P. 2010, arXiv:1007.3581

\bibitem[Kulkarni et al.(2005)]{Kulkarni05}
Kulkarni, V. P., Fall, S., M., Lauroesch, J. T., York, D. G. \& Welty, D. E. 2005, \apj, 618, 68

\bibitem[Lehner et al.(2004)]{Lehner04}
Lehner, N., Wakker, B. P. \& Savage, B. D. 2004, \apj, 615, 767

\bibitem[Lidz et al.(2011)]{Lidzetal11}
Lidz, A. et al. 2011, arXiv.org:1104.4800

\bibitem[Malaney \& Chaboyer(1996)]{Malaney96}
Malaney, R. A., \& Chaboyer, B. 1996, ApJ, 462, 57

\bibitem[Malhotra et al.(2001)]{Malhotra01}
Malhotra, S., et al. 2001, \apj, 561, 766


\bibitem[Nagamine et al.(2006)]{Nagamine06}
Nagamine, K., Wolfe, A. M. \& Hernquist, L. 2006, \apj, 647, 60

\bibitem[Naylor et al.(2003)]{Naylor03}
Naylor, B. J., et al. 2003, in Society of Photo-Optical Instrumentation Engineers (SPIE) Conference, Vol. 4855, Society of Photo-Optical Instrumentation Engineers (SPIE) Conference Series, ed. T. G. Phillips \& J. Zmuidzinas, 239–248

\bibitem[Oberst et al.(2006)]{Oberst06}
Oberst, T. E., et al. 2006, \apjl, 652, L125

\bibitem[Obreschkow et al.(2009a)]{Obreschkow09a}
Obreschkow, D., Croton, D., De Lucia, G., Khochfar, S. \& Rawlings, S. 2009a, \apj, 698, 1467

\bibitem[Obreschkow et al.(2009b)]{Obreschkow09b}
Obreschkow, D., Klockner, H.-R., Heywood, I., Levrier, F. \& Rawlings, S. 2009b, \apj, 703, 1890


\bibitem[Osterbrock(1989)]{Osterbrock89}
Osterbrock, D. E. 1989, Astrophysics of Gaseous Nebulae and Active Galactic Nuclei (Mill Valley: Univ. Science Books)

\bibitem[Pei et al.(1995)]{Pei95}
Pei, Y. C., \& Fall, S. M. 1995, \apj, 454, 69

\bibitem[Pei et al.(1999)]{Pei99}
Pei, Y. C., Fall, S. M., \& Hauser, M. G. 1999, \apj, 522, 604

\bibitem[Santos et al.(2008)]{Santos08}
Santos, M. G., et al. 2008, \apj, 689, 1

\bibitem[Santos et al.(2010)]{Santos10}
Santos, M. G., Ferramacho, L., Silva, M. B., Amblard, A. \& Cooray, A. 2010, \mnras, 406, 2421

\bibitem[Santos et al.(2011)]{Santos11}
Santos, M. G., Silva, M. B., Pritchard, J. R., Cen, R. \& Cooray, A. 2011, A\&A, 527, A93

\bibitem[Savaglio(1997)]{Savaglio97}
Savaglio, S. 1997, seim. proc. 73

\bibitem[Sheth \& Tormen(1999)]{Sheth99}
Sheth, R. K. \& Tormen, G. 1999, \mnras, 308, 119

\bibitem[Smith et al.(2003)]{Smith03}
Smith, R. E., et al. 2003, \mnras, 341, 1311

\bibitem[Spitzer(1978)]{Spitzer78}
Spitzer, L. J. 1978, Physical Processes in the Interstellar Medium (New York: Wiley)

\bibitem[Springel et al.(2005)]{Springel05}
Springel, V., et al. 2005, Nature, 435, 629

\bibitem[Stacey et al.(1991)]{Stacey91}
Stacey, G. J., Geis, N., Genzel, R., Lugten, J. B., Poglitsch, A., Sternberg, A. \& Townes, C. H. 1991, \apj, 373, 423

\bibitem[Stacey et al.(2010)]{Stacey10}
Stacey, G. J., et al. 2010, \apj, 724, 957

\bibitem[Suginohara et al.(1999)]{Suginohara99}
Suginohara, M., Suginohara, T. \& Spergel, N., 1999, \apj, 512, 547

\bibitem[Tayal(2008)]{Tayal08}
Tayal, S. S. 2008, A\&A., 486, 629

\bibitem[Tielens \& Hollenbach(1985)]{Tielens85}
Tielens, A. G. G. M. \& Hollenbach, D. 1985, \apj, 291,722

\bibitem[Visbal \& Loeb(2010)]{Visbal10}
Visbal, E. \& Loeb, A. 2010, JCAP, 11, 016

\bibitem[Walter et al.(2009)]{Walter09}
Walter, F., Riechers, D., Cox, P., Neri, R., Carilli, C., Bertoldi, F., Weiss, A. \& Maiolino, R. 2009, Nature, 457, 699

\bibitem[Wolfire et al.(1995)]{Wolfire95}
Wolfire, M. G., et al. 1995, \apj, 443, 152

\bibitem[Wouthuysen(1952)]{Wouthuysen52}
Wouthuysen, S. A. 1952, \aj, 57, 31

\bibitem[Wright et al.(1991)]{Wright91}
Wright, E. L., et al. 1991, \apj, 381, 200

\end{thebibliography}
\end{document}